\documentclass{aa}
\usepackage{graphics,latexsym,amssymb,times,psfig}

\def\gsim{ \lower .75ex \hbox{$\sim$} \llap{\raise .27ex \hbox{$>$}} } 
\def\lsim{ \lower .75ex\hbox{$\sim$} \llap{\raise .27ex \hbox{$<$}} } 

\begin{document}
\newcommand {\sax} {{\it Beppo}SAX }
\newcommand {\gpeak} {$\gamma_{\rm peak}$ }
\newcommand {\als} {$\alpha_1$}
\newcommand {\alh} {$\alpha_2$}
\newcommand {\arx} {$\alpha_{\rm rx}$ }
\newcommand {\aro} {$\alpha_{\rm ro}$ }
\newcommand {\aox} {$\alpha_{\rm ox}$ }

\title{TeV candidate BL Lac objects}


\author{Luigi Costamante\inst{1,2} and Gabriele Ghisellini \inst{2} }

\offprints{L. Costamante; costa@merate.mi.astro.it}
\institute{
Universit\`a Statale di Milano, via Celoria 16, I--20133 Milano,  Italy.
\and Osservatorio Astronomico di Brera, via Bianchi 46, I--23807 Merate, Italy.
}

\date{Received 2001, Accepted 2001}
 
\titlerunning{TeV candidate BL Lac objects}
\authorrunning{L. Costamante \& G. Ghisellini}

\abstract{
The TeV emission of low power BL Lac objects has been established
by the detection of an handful of them.
The knowledge of the level of the TeV emission and its spectrum can shed
light on the particle acceleration mechanisms,
and it is especially important to assess the still uncertain
level of the far infrared background radiation, which can absorb 
the TeV photons through photon--photon interactions.
In view of these implications, it is necessary to
enlarge the number of TeV detected sources, and to find them at 
different redshifts.
To this aim, we propose a general and simple criterium to select
the best TeV candidates, and produce a list of them with flux estimates
above 40 GeV, 300 GeV and 1 TeV. 
\keywords{Galaxies: jets --- Galaxies: nuclei --- BL Lacertae objects: general 
--- Radio continuum: galaxies ---  X--rays: galaxies}
}
\maketitle

\section{Introduction}

The discovery that blazars (Flat Spectrum Radio Quasars and BL Lac objects)
are very strong $\gamma$--ray emitters have renewed the interest about them,
and opened new perspectives in the comprehension of the physics
of these objects.

The observations by EGRET (Hartmann et al. 1999), 
onboard the {\it Compton Gamma Ray Observatory},
led to the discovery that blazars emit most of their 
power in the $\gamma$--ray band, and that their Spectral Energy 
Distribution (SED) is characterized by two broad peaks, now commonly
(but not unanimously), interpreted as due 
to synchrotron and inverse Compton (IC) radiation, respectively
(e.g. Maraschi et al. 1992; Dermer et al. 1992;
Sikora et al. 1994; Ghisellini \& Madau 1996;
but see Mannheim 1993; Rachen 1999; Muecke \& Protheroe 2000; Aharonian 2000
for a different interpretation).
For the first time we were able to study their entire SED 
in a comprehensive way, finding differences among subclasses
of blazars about the frequency location of the two broad peaks,
their relative luminosity and variability behaviors in different bands.
Considering the EGRET sources and three complete blazar samples,
it was found a correlation between the location of the two broad
peaks and the observed bolometric luminosity
(Fossati et al. 1998, hereafter F98; Ghisellini et al. 1998).
Blazars seem to form a well defined sequence, with low powerful 
objects having both peaks at a similar level of luminosity, and 
located at higher frequencies than in more powerful objects, 
in which the IC peak dominates the emission.
In the BL Lac class,  
the first kind of sources were named High frequency Peak BL Lacs
(or HBL for short) by  Padovani \& Giommi (1995), while the latter 
subclass was called Low frequency Peaked BL Lacs (LBL).
Recent observations of high redshift ($z>4$) blazars (Fabian et al. 
2001a, 2001b; Celotti 2001), 
and of low power BL Lac objects (Costamante et al. 2001a, 2001b) 
have extended the blazar sequence at both ends, resulting in agreement with the 
original trend. 

At TeV energies, the detection and study of blazar objects 
by ground based Cherenkov telescopes have been limited
up to now to few HBL sources 
(Mkn 501, Mkn 421, PKS 2155--304, 1ES 2344+514, see Catanese \& Weekes 1999),
though disclosing new and 
fundamental aspects of the blazar behavior. 
These observations monitor the behavior of the most
energetic electrons of the source, thus shedding light on the 
acceleration mechanism working at the most extreme conditions.
The very rapid variability observed at these energies
(Mkn 421 doubled its TeV flux in less than 20 minutes, 
see Gaidos et al. 1996, Catanese \& Weekes  1999), coupled with the 
requirement that the source must be transparent with respect 
to the photon--photon process, tightly constrains the physical 
parameters, such as the source size and its beaming factor.
In addition, Mkn 421 showed a tight correlation between the emission
in the X--ray and TeV bands (Maraschi et al. 1999, Takahashi et al. 2000, 
Krawczynski et al. 2001),
implying that the radiation produced
in the two bands is co-spatial and produced by the same electrons:
this is of crucial importance to constrain any emission model.
%
%

The strong connection between the TeV and X-ray emission was also clearly evident
during the 1997 flare of Mkn 501, when this source was observed by the 
X-ray satellite {\it Beppo}SAX in an extreme spectral state, 
with a synchrotron peak frequency close to 100 keV or even more
(Pian et al. 1998). 
Mkn 501 was found to have
increased at least tenfold its luminosity, with most of it
radiated at high X--ray energies.
At the same time, the source underwent a major flare in the TeV band
(Catanese et al. 1997b; Aharonian et al. 1997; Protheroe et al. 1997;
Djannati--Ata\"{\i} et al. 1999), and continued to be active 
(and well visible) in the TeV band for several months.
This dramatic behavior can be explained by a synchrotron 
inverse Compton model, taking into account the effects introduced 
by the Klein Nishina scattering cross section and the constraints 
posed by the transparency of the source with respect to photon--photon 
collisions producing electron--positron pairs.
It is not clear if the simultaneous variations in the X--ray and TeV bands 
can be completely accounted for by a simple one--zone homogeneous synchrotron 
self--Compton  model (see Tavecchio et al. 2001), or 
if we need some extra and more quiescent source of IR--optical
photons (i.e. Ghisellini 1999).

TeV observations of BL Lacs are also particularly interesting 
because, being the only extragalactic sources known to emit at these
energies, allow an independent estimate of the extragalactic 
IR background (IRB).
Direct measurements of the IRB are affected (up to now) by relatively
large uncertainties, due to the heavy contamination of foreground objects
(for a review see Hauser \& Dwek 2001).
Because TeV photons can be absorbed by IR photons
for the pair production mechanism, the analysis/study of 
high energy spectra from sources at different redshifts
gives an independent measure of the IRB level (Stecker et al. 1992).

These kind of studies are just started, and the first conclusions
are based on the two most observed TeV BL Lacs, namely Mkn 421 and Mkn 501.
The main uncertainty here is the knowledge of the primary
blazar spectrum, which could have an intrinsic cut--off at high energies,
or could be affected by absorption due to IR photons produced locally.
A first and preliminary confirmation of the IRB absorption has been
obtained comparing the spectra of Mkn 421 and Mkn 501, 
showing a cut--off in the power--law spectra at approximately the same
energies (Krennrich et al. 2001), as expected since they have similar redshifts.

The direct measurements of the IRB flux 
lead to predict quite a strong absorption at TeV energies. 
If true, this in turn would imply an unusual primary spectrum above
$\sim 1$ TeV, which must have an excess above the extrapolation from
lower energies, leading to the so--called IRB--TeV puzzle 
(see e.g. Protheroe \& Meyer 2000; Aharonian, Timokhin \& Plyasheshnikov, 2001; 
Berezinsky 2001)
{\footnote{It has been proposed that this ``puzzle" could be solved by
quantum--gravity theories predicting the breaking of Lorentz invariance:
as a consequence there should be a modification in the energy
threshold for the $\gamma$--$\gamma$ $\to$ $e^\pm$ process, explaining
why TeV photons are not heavily absorbed by the IRB (Amelino--Camelia
\& Piran 2001).}.

However, to draw unambiguous conclusions, we need an ensemble of sources 
located at different redshifts and a detailed knowledge of the
X--ray flux and spectrum: being produced by the same electrons 
(in synchrotron inverse Compton models), this would help in predicting
the shape of the TeV emission.

To this aim we consider in this paper several published samples of
bright BL Lac objects, for a total of 246 different objects,
and propose a simple and handy tool to identify and select the most 
promising candidates.
The main point we emphasize concerns the requirement of {\it both}
high energy electrons {\it and} sufficient seed photons to originate
the TeV emission. 
We therefore consider as best candidates those BL Lac objects
having not only their synchrotron peak located at high energies,
but also having sufficient radio--through--optical flux.

We therefore expand the work first done by Stecker et al. (1996),
concerning only the Einstein Slew survey sample of BL Lac objects,
both by considering other samples and also by introducing a different,
albeit still simple, criterium to identify the best candidates.


\begin{figure*} 
\psfig{figure=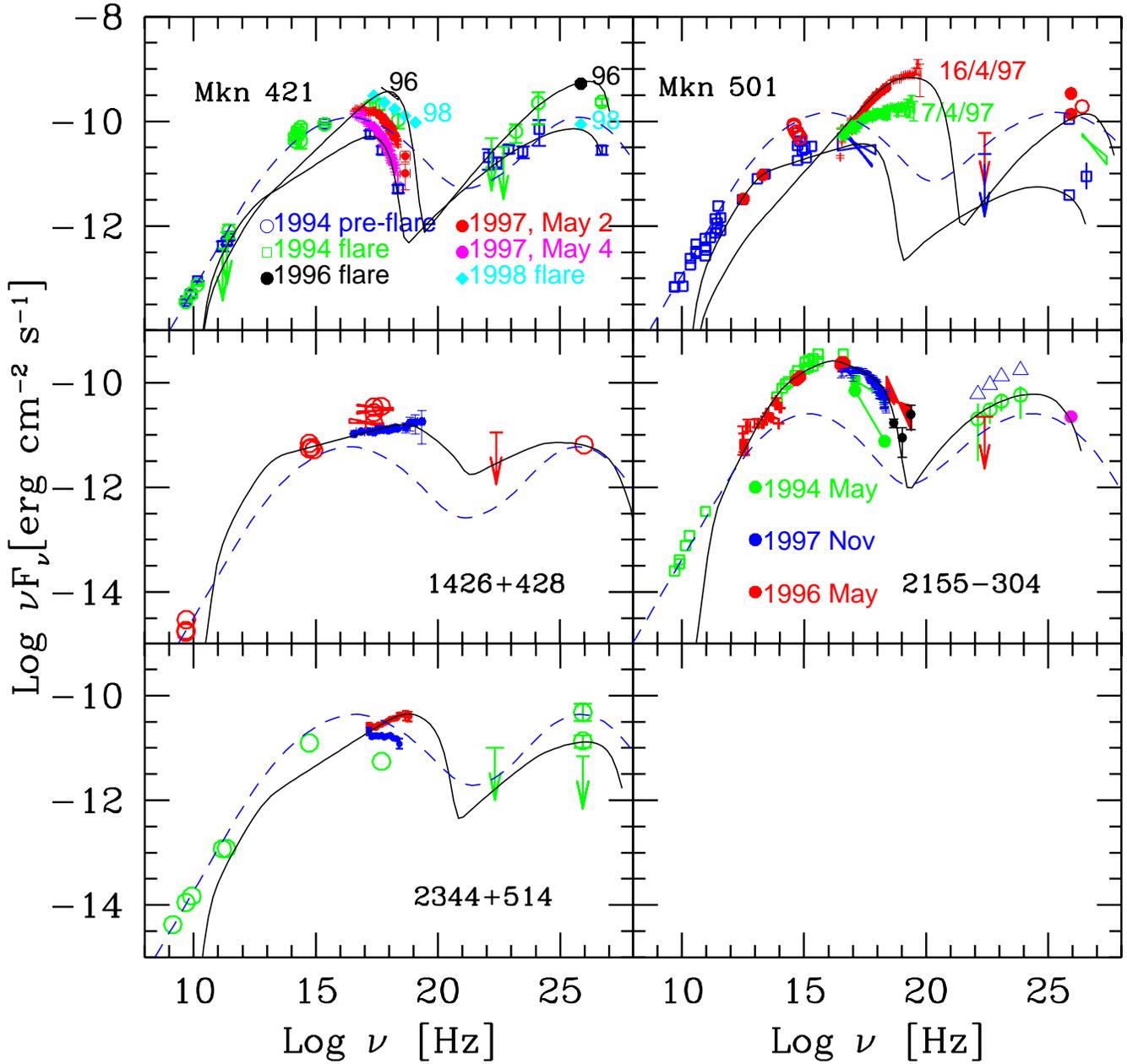,width=18cm,height=18cm}
\vskip -0.8 true cm
\caption{Spectral energy distributions of the 5 BL Lac objects
already detected at TeV energies.
The solid and dashed lines refer to the SSC model described in Section 3.2
and to the SED constructed using the parameterization described
in Fossati et al. 1998, with the modification described in this paper.
The input parameters used for the SSC models are given in
Ghisellini, Celotti \& Costamante (2001).
} \label{tevdet}
\end{figure*}

\section{BL Lac objects detected at TeV energies}

At present, only few blazars have been detected by Cherenkov 
telescopes. 
In Table  \ref{tevref} we report the basic properties, fluxes and detection 
thresholds for those objects detected at least once above $5\sigma$
(see Catanese \& Weekes 1999 and references therein).
We have also added 1ES 1426+428, 
whose detection was reported  
at about the 4.5$\sigma$ level by WHIPPLE (Horan D. et al., 2000).
This was confirmed by CAT (Djannati--Atai, priv. comm.), by WHIPPLE 
again in 2001 (Horan D. et al., 2001), and  
by HEGRA (Aharonian 2001b). The latter two detections were at
the 5$\sigma$ level.


Of these objects however, only Mkn 421 and Mkn 501 have been firmly
confirmed by repeated detections by many Cherenkov telescopes,
and since 1995 they  have been extensively monitored.
Thanks to their relatively high fluxes and repeated activity states,
these are also the only two sources with spectral informations
up to $\sim17$  and 20 TeV (Krennrich et al. 2001). 

Two other objects have been reported to be TeV emitters, but with a 
somewhat lower significance. 
1ES 1959+650 was observed in 1998 by the UTAH Seven Telescope Array, 
and found with an excess of $3.9\sigma$, reaching in two periods  
$\sim 5\sigma$, but no fluxes were reported (Nishiyama et al. 1998). 
3C 66A was instead detected once at $\sim 4.2\sigma$ by the Crimean 
Astrophysical Observatory in 1996 (Neshpor et al. 1998), with a detection 
threshold at 0.9 TeV. 
If confirmed, this source could be extremely important, since due to the 
high redshift ($z$=0.444), its TeV emission should be suppressed 
according to the actual estimates on the $\gamma$--$\gamma$ $\to$ $e^\pm$ 
optical depth due to the IRB (Stecker et al. 1999).

\begin{table*}
\begin{center}
\begin{tabular}{lllllllll}
\hline
Source  &$z$ &$F_{\rm 5 GHz}$  &$F_{\rm 5500 A}$  &$F_{\rm 1 keV}$  &$F^1_{\rm TeV}$ &$E_{\rm th}$ & Signif. &Tel.\\
        &    &Jy               &mJy               &$\mu$Jy          &        &  TeV &  &  \\
\hline
Mkn 421        &0.031  &  0.722  & 9.75  & 9.98  & 1--80  &  0.3  & conf.  &  WHIPPLE, CAT, HEGRA \\
1ES 1426$+$428 &0.129  &  0.038  & 1.56  & 6.83  &  0.9   & 0.4  & $4.5-5\sigma$ &  WHIPPLE, (CAT), HEGRA  \\
Mkn 501        &0.034  &  1.371  &10.31  & 9.44  & 0.8--49  & 0.3  &  conf.  & WHIPPLE, CAT, HEGRA \\
PKS 2155$-$304 &0.116  &  0.310  &22.34  & 14.6  & 4.2     & 0.3  & $6.8\sigma$ & Durham Mark6  \\
1ES 2344$+$514 &0.044  &  0.215  & 3.54  & 2.91  & 1.7    & 0.35  & $5.2\sigma$ & WHIPPLE  \\
\hline
\end{tabular}
\caption{ Data for the BL Lac objects detected in the TeV band.
Significances are listed only for unconfirmed sources. Data from
Catanese \& Weekes  (1999) and Krennrich et al. (2001), and references therein.
1) integrated flux in units of $10^{-11}$ photons cm$^{-2}$ s$^{-1}$. } 
\label{tevref}
\end{center}
\end{table*}

\section{The Synchrotron self Compton process and TeV emission}

The BL Lac objects so far detected at TeV energies are relatively low
powerful blazars, with no broad emission lines and no signs
of thermal emission (i.e. the blue bump) produced by the accretion disk.
This suggests and favors a synchrotron self--Compton (SSC) origin
of the TeV flux.
However this may be an over--simplification, since, besides
the jet region containing the energetic electrons responsible
for the high energy emission, other sites, both in the jet and
externally to it (e.g. a molecular torus, a thin scattering plasma 
surrounding the jet, or the walls of the jet itself), may be important 
in producing  the soft seed photons to be scattered at high energies.
However, a simple one--zone and homogeneous SSC model can account
for the observed SED of all the TeV BL Lacs, and due to its
simplicity it is the basic framework we will use in the following.
It is also the model assumed by the previous work
on this issue, namely the paper by Stecker et al. (1996),
which we discuss below, before introducing our views.

\subsection{The Stecker et al. (1996) scenario}

Based on the SSC framework, 
Stecker, De Jager \& Salamon (1996) have used simple scaling arguments
to predict the $\gamma$-ray fluxes for HBL objects, and so
to select good candidates for TeV emission.
According to the SSC model, the inverse Compton component has a spectrum 
which is similar to the synchrotron one (both roughly parabolic on a logarithmic 
$\nu F_{\nu}$ plot, Macomb et al. 1995, F98), 
but upshifted by $\sim\gamma^2_{\rm peak}$ (in the Thomson regime), 
where $\gamma_{\rm peak}$ is the  Lorentz factor of the electrons emitting 
at the peak.  
Using as ``template" the SED of Mkn 421, at the time the only HBL source
detected both at GeV (EGRET) and TeV (WHIPPLE) energies 
(i.e. with informations on both sides of the Compton peak),
they found an upshifting factor of $\sim10^9$, and a $L_{\rm C}/L_{\rm syn}\sim1$.
Assuming then for simplicity that all HBL objects have the same properties
as those found for Mkn 421, the found upshifting factor allowed to derive
the following scaling law:
\begin{equation}
\frac{\nu_{\rm o}F_{\rm o}}{L_{\rm syn}}\simeq\frac{\nu_{\rm GeV}F_{\rm GeV}}{L_{\rm C}}
\;\;{\rm and}\;\;
\frac{\nu_xF_x}{L_{\rm syn}}\simeq\frac{\nu_{\rm TeV}F_{\rm TeV}}{L_{\rm C}},
\end{equation}
where $F_{\rm o}$ is the monochromatic flux at the optical frequency $\nu_{\rm o}$.
Given $L_{\rm C}/L_{\rm syn}\sim1$, a direct relation for the energy fluxes is obtained,
in the GeV and TeV ranges:
\begin{equation}
\nu_{\rm GeV}F_{\rm GeV} \sim \nu_{\rm o}F_{\rm o} \;{\rm and}\;
\nu_{\rm TeV}F_{\rm TeV} \sim \nu_xF_x
\end{equation}
Quantitative estimates on the integral fluxes were then made using
the average spectral index for BL Lacs between 0.1 and 10 GeV ($\alpha=0.8$),
and the Mkn 421 slope above 0.3 TeV ($\alpha=1.2$).

\begin{figure} 
\vskip -0.7 true cm
\psfig{figure=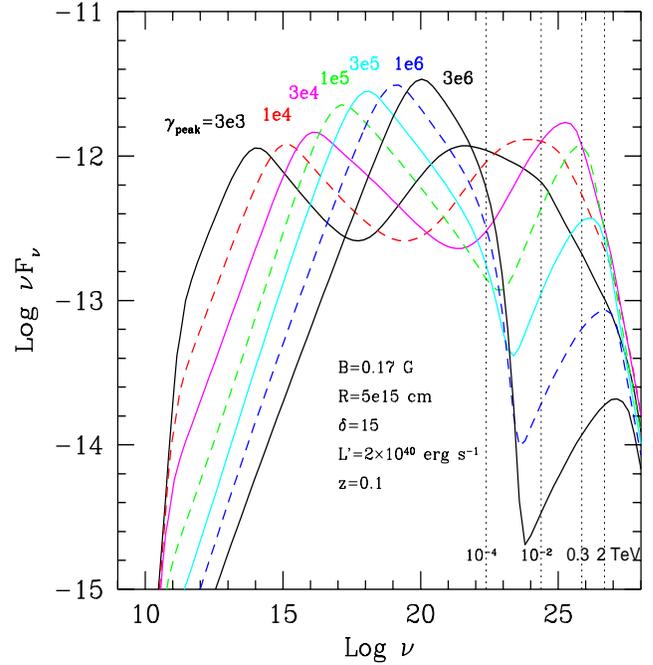,width=9.5cm,height=10.5cm}
\vskip -0.8 true cm
\caption{SEDs calculated with a simple one--zone SSC model as described in Section 3.2,
in which $\gamma_{\rm peak}$ increases (as labeled).
All other parameters are kept constant.
}
\end{figure} 

\begin{figure} 
\psfig{figure=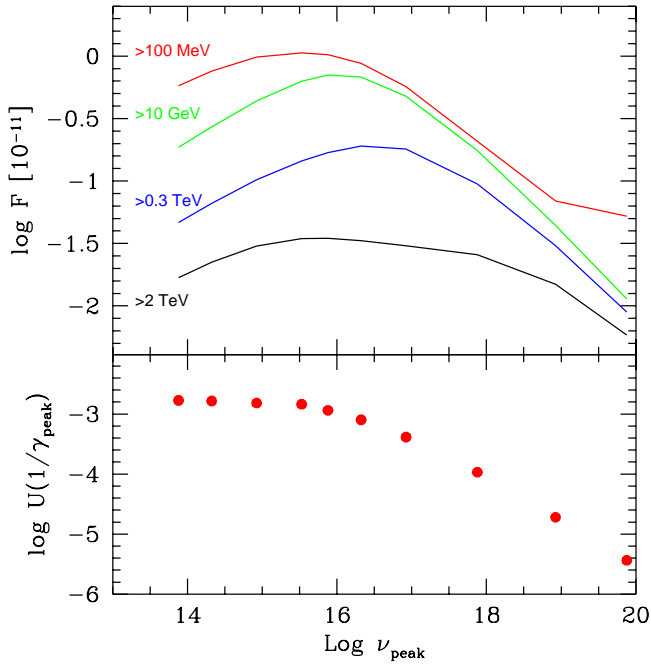,width=9.5cm,height=9.5cm}
\vskip -0.5 true cm
\caption{Top panel shows the flux integrated above
the indicated energies corresponding to the spectra of Fig. 2.
$\nu_{\rm peak}$ is the synchrotron peak frequency of the SEDs in Fig. 2.  
The bottom panel shows the radiation energy density
available for scattering in the Thomson regime with
electrons of energy $\gamma_{\rm peak} m_e c^2$.
}
\end{figure}

In this scenario, therefore, the main selection criterium is the 
X--ray flux level: the energy flux around ~1--2 keV gives directly the expected 
level of the emission roughly around 0.2--0.3 TeV.
In this context, sources with the highest 
synchrotron $\nu_{\rm peak}$ are expected 
to emit more in the TeV band, 
since the Compton peak progressively moves into the TeV window.

\subsection{Energetic electrons and seed photons}

The two ingredients for the formation of a strong TeV emission
by the synchrotron self Compton process are the density of electrons
energetic enough to produce TeV photons and the density of seed
photons to be scattered.
For not too large synchrotron peak frequencies, the 
scattering process between photons at the synchrotron peak and
the electron producing the peak itself is in the Thomson regime.
In this case the relevant electrons can scatter the bulk of the
synchrotron photons to high energies.

For larger synchrotron peak frequencies (and assuming the same value of the 
magnetic field) we have more energetics electrons in principle capable
to produce photons of larger energies.
However in this case the Klein--Nishina decline of the scattering
cross section is important, disfavoring those scatterings between 
the relevant electrons and the synchrotron photons lying at the peak.
In this case the scattering process producing TeV photons is produced
by seed photons of lower energies, with a corresponding reduced
synchrotron energy density.
In other words, not all the synchrotron photons are used to form
the Compton spectrum, but only the photons of energy (in the comoving,
primed, frame) $h\nu^\prime  <  m_e c^2/\gamma$
can efficiently (i.e. they scatter in the Thomson regime) 
contribute to the Compton emission.
There is then a trade--off: to produce large TeV fluxes we need
very energetic electrons, corresponding to very large $\nu_{\rm peak}$,
but we also need enough seed photons, and this requires not
extreme values of $\nu_{\rm peak}$.

To better illustrate this case we show in Fig. 2 a sequence of
spectra derived by a standard SSC model in which all parameters but 
$\gamma_{\rm peak}$ are kept constant.
In this model it is assumed that relativistic
electrons between $\gamma_1$ and $\gamma_2$ are injected
at a constant rate $Q(\gamma) \propto \gamma^{-2.5}$ [cm$^{-3}$ s$^{-1}$],
throughout a spherical source of radius $R$,
magnetic field $B$ and beaming factor $\delta$.
The steady state particle distribution is found through 
the continuity equation, accounting for synchrotron and inverse Compton
radiative losses, electron positron pair production
and Klein Nishina effects. 
Particles are assumed not to escape the source.
In this case $\gamma_{\rm peak}$ is always coincident with $\gamma_1$
(see Ghisellini et al. 1998 for more details of the model).

In Fig. 3 we report the flux integrated above a given frequency
as a function of $\nu_{\rm peak}$, corresponding to the spectra
shown in Fig. 2.
In the bottom panel we also show the comoving radiation energy density
which is available for scattering in the Thomson limit. From 
both Fig. 2 and Fig. 3 we can see that the level of the TeV 
emission initially increases for increasing $\gamma_{\rm peak}$, 
since in this case the scattering of the bulk of the synchrotron photons 
occurs in the Thomson regime, and the increase in $\gamma_{\rm peak}$ 
results in more photons reaching TeV energies.
This trend is reversed when $h \nu^\prime_{\rm peak} \ge m_ec^2 
/\gamma_{\rm peak}$, e.g. when
\begin{equation}
\nu_{\rm peak}  \, \ge \, 3.8\times 10^{15} B^{1/3} {\delta \over 1+z}
\quad {\rm Hz}
\end{equation}
where $\nu_{\rm peak} = (4.3)\nu_{\rm L}\gamma^2_{\rm peak}
\delta/(1+z)$, and $\nu_{\rm L}=eB/(2\pi m_e c)$ is the Larmor frequency.
The increase in the synchrotron luminosity for increasing
$\nu_{\rm peak}$ is due to the corresponding decrease in the 
self Compton component: since all the injected power is assumed
to be radiated, the sum of the synchrotron and the self
Compton luminosities must be constant.

We conclude that very extreme BL Lacs, with very large values
of $\nu_{\rm peak}$, are not necessarily the best candidates
to be strong TeV emitters since they can scatter relatively fewer
photons.
The best TeV BL Lac candidates should be the one with {\it both}
a large $\nu_{\rm peak}$ {\it and} a sufficiently strong
soft seed IR photon emission.

\section{TeV candidate BL Lacs}

\subsection{The Samples}

In order to select new candidates for TeV emission, we have assembled a 
catalogue of BL Lac objects using several published BL Lac samples, 
for which informations in all the three energy bands 
(radio, optical and X--ray) were available.
We considered the Slew Survey Sample (Perlman et al. 1996), 
the Einstein Medium Sensitivity Survey (EMSS, Rector et al. 2000), 
the ROSAT All Sky Survey BL Lac sample (RASS, Bade et al. 1998),
the ROSAT All Sky Survey -- Green Bank sample (RGB, 
Laurent--Muehleisen et al. 1999), the EXOSAT archive BL Lac 
catalogue (Giommi et al. 1990, Sambruna et al. 1994) and 
the 1 Jy BL Lac sample (Stickel et al. 1993, Urry et al. 1996).
We have also added  all the HBL objects in Donato et al. (2001), 
who gives a list of all known blazars detected in X--rays for 
which a measure of the spectral index was available.

The sources listed in more than one sample have been considered only once,
and attributed with the following order: Slew, EMSS, RASS, RGB, EXOSAT, 1 Jy samples
and those in the Donato et al. compilation  
(i.e. the EMSS sources in Fig. \ref{fxfr} and \ref{fxfo} 
are those not already in the Slew sample, the RASS
sources are those not included in the EMSS and Slew samples, and so on).  
In this way we obtained an ensemble of 246 different objects. 
The flux informations used are those reported in the respective catalogue 
papers (see each reference).
Because the data were often provided in different formats, 
we have uniformed all the fluxes to monochromatic fluxes at 5 GHz, 5500\AA~ 
and 1 keV, using the same spectral information adopted in the compilation 
of each sample, for consistency.
In the radio band, the RASS sample reported the fluxes at 1.4 GHz along with  
the radio spectral index, so we have calculated the flux at 5 GHz assuming
a power--law spectrum and the reported $\alpha_{\rm R}$. 
In the X--ray band we calculated the fluxes at 1 keV 
from integrated fluxes (RASS and RGB) or fluxes at 2 keV (Slew, EMSS) using the   
X--ray spectral indices there reported. 
The optical fluxes were calculated from the tabulated magnitudes, dereddened
with the $A_{\rm B}$ values obtained from the NED database (Burstein \& Heiles 1982). 
When the optical flux was not at 5500\AA~(V filter), we obtained it 
extrapolating from the flux at the different effective wavelength, 
assuming a power law spectrum of slope $\alpha_{\rm opt}=1$.
The NED database was also used to check the redshift for all sources, 
65 of which do not have a reported value.

For Fig. \ref{fxfr} and \ref{fxfo}, the fluxes have been K--corrected 
with the respective catalogue  spectral indices, assuming power--law spectra.
When the slope was not reported, we used $\alpha_{\rm R=0}$ and $\alpha_{\rm opt}=1$
for the K--correction of the radio and optical fluxes, respectively.
The X--ray flux was K--corrected using the reported spectral indices.
For the K--correction in case of unknown redshift we used $z=0.2$. 

\begin{figure*} 
\vskip -1 true cm
\psfig{figure=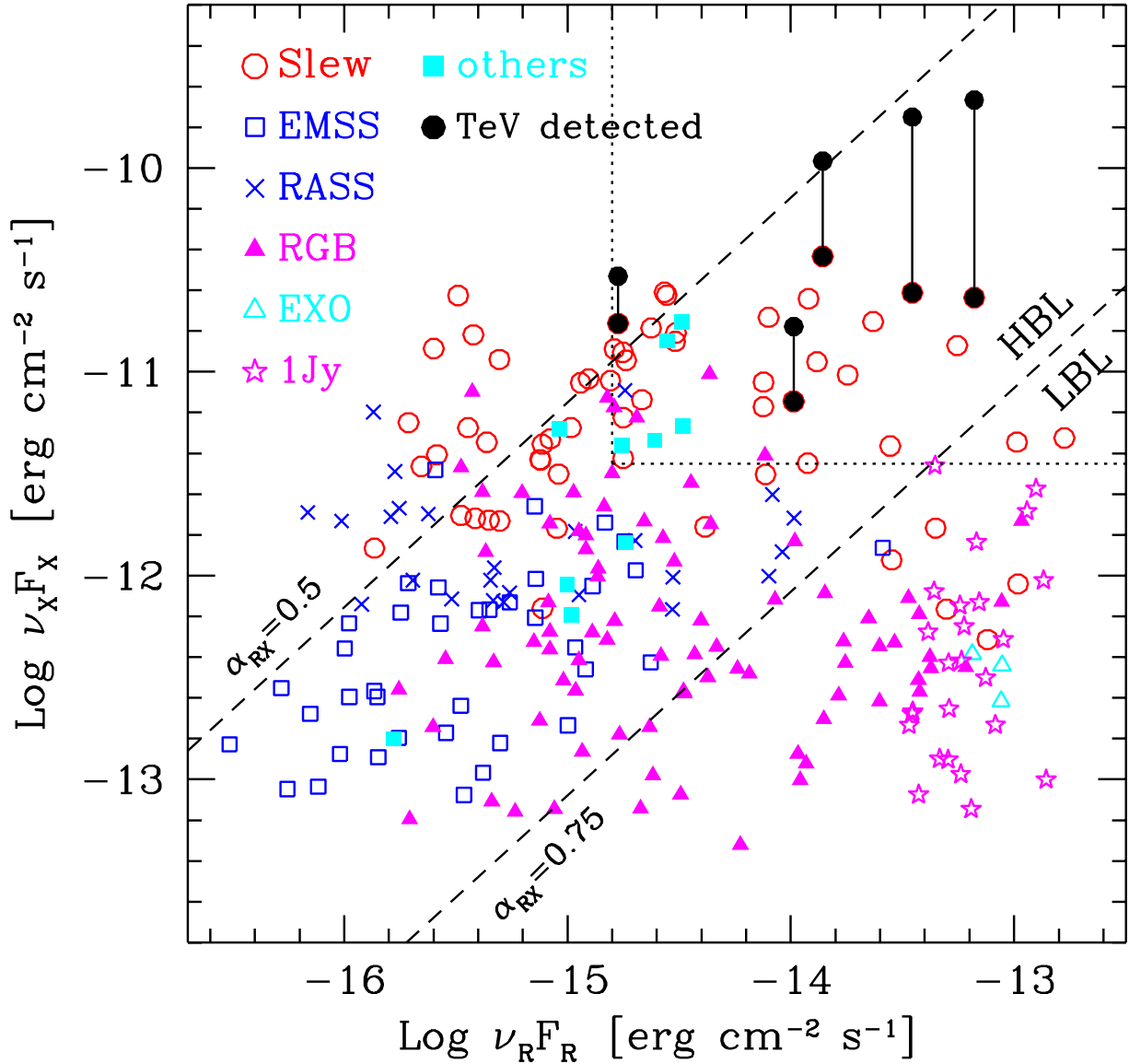,width=18.5cm,height=18.5cm}
\vskip -1 true cm
\caption{BL Lac objects in the radio (5 GHz) and X--ray (1 keV)
$\nu F(\nu)$ plane.
Sources belonging to different samples have
different symbols, as labeled.
The objects marked with filled circles are those already detected 
at TeV energies (from left to right, 1ES 1426+428, 1ES 2344+514, 
PKS 2155--304, Mkn 421, Mkn 501).
Note that for these sources we have plotted two different X--ray states,
connected by the vertical segment.
The dotted lines delimiting the rectangle are at 
$F_{\rm x}=1.46 \mu$Jy and $F_{\rm R}=31.6$ mJy.} 
\label{fxfr}
\end{figure*} 

\begin{figure*} 
\vskip -1 true cm
\psfig{figure=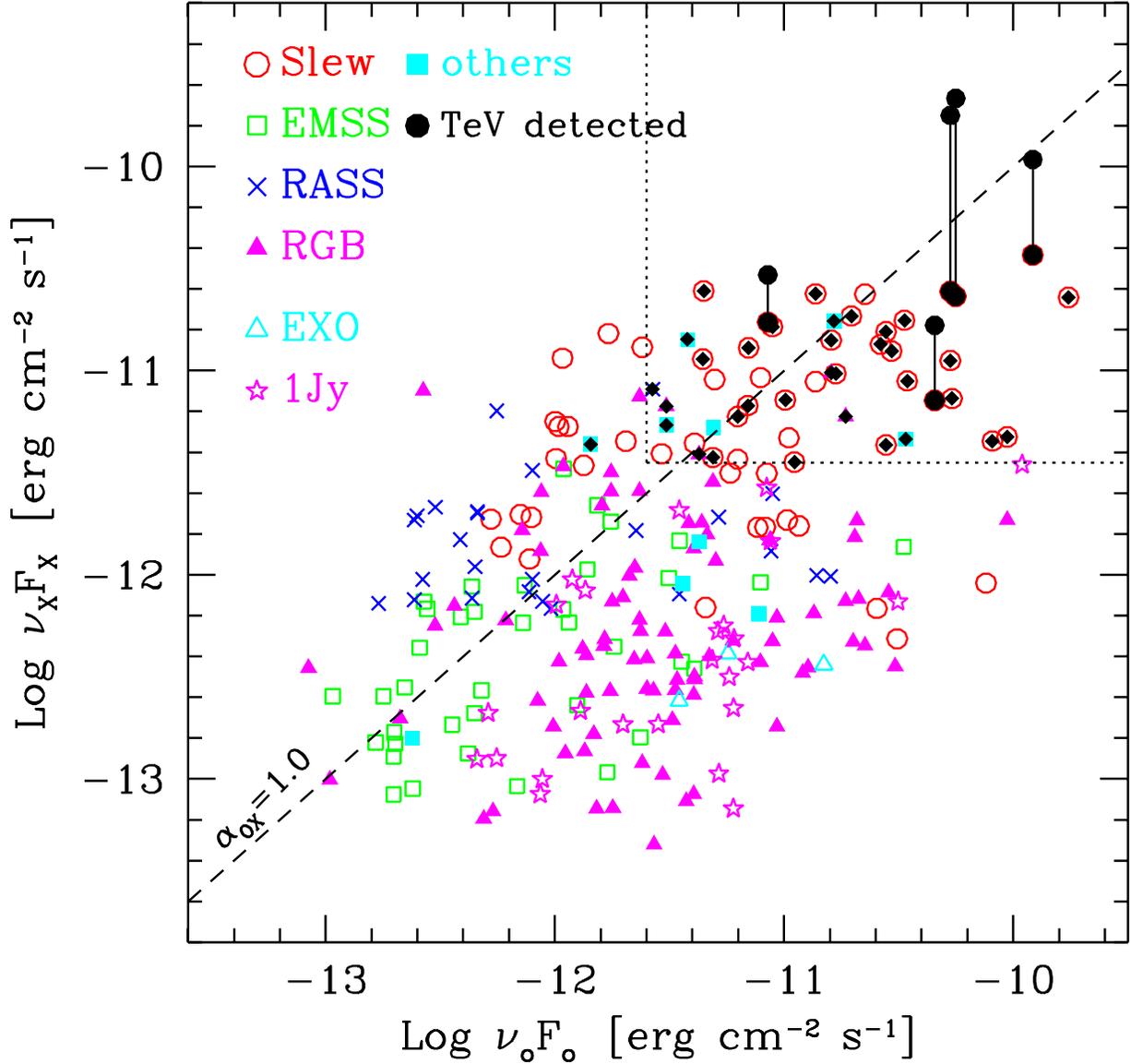,width=18.5cm,height=18.5cm}
\vskip -1 true cm
\caption{BL Lac objects in the optical (5500 \AA) and X--ray (1 keV)
$\nu F(\nu)$ plane.
Sources belonging to different samples have different symbols, as labeled.
The objects marked with filled circles are those already detected 
at TeV energies (from left to right, 1ES 1426+428, 1ES 2344+514, Mkn 421, 
Mkn 501 and PKS 2155--304).
Note that for these sources we have plotted two different X--ray states,
connected by the vertical segment.
The dotted lines delimiting the rectangle are at $F_{\rm x}=1.46 \mu$Jy and $F_{\rm o}=0.49$ mJy.
The objects marked also with a black diamond are those inside the rectangle of 
the $\nu_{\rm x} F_{\rm x}$--$\nu_{\rm R} F_{\rm R}$ plot (Fig. \ref{fxfr}).
} \label{fxfo}
\end{figure*}

\begin{table*}
\begin{center}
\begin{tabular}{|llllllllllll|}
\hline
 Name     &S &Ex &Ra & R &1J &oth. &$z$ &$F_{\rm 5~GHz}$ &$F_{5500{\tiny \AA}}$ & $F_{\rm 1 keV}$  &Ref \\              
          &  &   &    &   &    &     &    &Jy              &mJy            &$\mu$Jy   &    \\   
\hline 
0033$+$595 &X  &  &   &X &  &  &0.086$^a$ &0.066 &2.95  &5.66    & Co01, SG99, Br97 \\
0120$+$340 &X  &  &   &X &  &  &0.272  &0.045   &1.23   &4.73    & Co01, BS94  \\   
0136$+$391 &   &  &X  &X &  &  & ---   &0.049   &3.41   &2.39    & Ba00, Br97\\
0214$+$517 &   &  &X  &X &  &  &0.049  &0.161   &0.39   &1.60    & Ne94  \\
0219$+$428 &X  &X &   &  &  &  &0.444  &0.806   &5.12   &1.57    & Pi93, Si91, Ro00, WW90, C97, Di96 \\
0229$+$200 &X  &  &   &  &  &  &0.139  &0.049   &9.93   &2.88    & Ah00\\
0323$+$022 &X  &X &   &X &  &  &0.147  &0.042   &0.81   &4.49    & Fo98, SG99, Be01, Ch99 \\
0414$+$009 &X  &X &   &X &  &  &0.287  &0.070   &0.82   &9.29    & Wo00, Ah00 \\
0548$-$322 &X  &X &   &  &  &  &0.069  &0.170   &3.61   &7.47    & Co01,  GAM95 , WW90, Ta95, Ti94, Ro99, Bk99 \\
0556$-$384 &   &  &   &  &  &  &0.034  &0.068   &0.56   &2.2     & SG96, GT96, Ci95 \\
0647$+$250 &X  &  &   &X &  &  & ---   &0.073   &5.12   &6.01    &SG99, GAM95, BS94, Ah00  \\
0806$+$524 &X  &  &   &X &  &  &0.138  &0.172   &6.33   &3.51    &SG99, GAM95, BS94   \\
0809$+$024 &   &  &   &  &  &X & ---   &0.071   &6.23   &1.85    &Ba00, Br97  \\  
0851$+$203 &X  &X &   &X &X &  &0.306  &1.740   &14.9   &1.70    &To96, Br89, IN88, Ge94, \\
           &   &  &   &  &  &  &       &        &       &        &E94, Li94, Sa94, Ma88, P01 \\  
1011$+$496 &X  &  &X  &X &  &  &0.200  &0.286   &2.04   &1.38    &Fo98   \\
1028$+$511 &X  &  &X  &X &  &  &0.361  &0.044   &1.29   &4.80    &  Fo98, Be01   \\
1101$-$232 &X  &  &   &  &  &  &0.186  &0.066   &2.52   &9.25    &Wo00, SG99, Ch99  \\
1114$+$202 &   &  &   &  &  &X &0.139 &0.074   &3.03   &7.31     &  Co01, Br97   \\ 
1133$+$704 &X  &X &X  &X &  &  &0.045 &0.274   &9.75   &4.56     &  IN88, La96, Pir00, Bk99\\ 
1136.5$+$6737& &  &X  &X &  &  &0.135  &0.040   &0.49   &3.40    & F96, Ba94\\
1215$+$303 &X  &X &X  &X &  &  &0.237  &0.445   &3.08   &3.7     & GAM95, Fo98, Te98    \\     
1218$+$304 &X  &X &X  &X &  &  &0.182  &0.056   &1.63   &6.39    & Co01, GAM95, Pi93, Sa94, Fo98, Pir00  \\
1417$+$257 &   &  &X  &X &  &  &0.237  &0.040   &0.56   &2.65    & Ba00, La96\\
1440$+$122 &X  &  &   &X &  &  &0.162  &0.041   &0.90  &1.47    & Ba00, Br95\\
1544$+$820 &X  &  &   &  &  &  & ---   &0.043   &1.15   &2.31    &GAM95, Fo98, Be01  \\
1553$+$113 &X  &  &   &X &  &  &0.360  &0.636   &6.15   &6.54    &GAM95, Ne94, Fa90, IN88, Be01 \\ 
1722$+$119 &   &X &X &X  &  &  &0.018  &0.088   &2.95   &4.02    &Be92, Pi93, GAM95, SG99, Sa94 \\ 
1727$+$502 &X  &X &  &X  &  &  &0.055  &0.159   &1.27   &2.73    &Fo98, Pi93, Ke95, Bk99   \\
1741$+$196 &X  &  &  &X  &  &  &0.084  &0.223   &1.86   &2.88    &Br97, Pir00   \\
1959$+$650 &X  &  &  &   &  &  &0.047  &0.252  &1.35   &9.29  & GAM95, SG99, F96, Be01, Bk99  \\ 
2005$-$489 &X  &X &  &   &X &  &0.071  &1.192   &4.84   &5.42    &Ta01, Ki99, Bk99    \\
2200$+$420 &X  &X &  &X  &X &  &0.069  &3.593   &17.27  &1.91    & St94, Br89, IN88, To96, Li94, Bk99, \\
          &   &  &   &  &  &  &       &        &       &         & Ca97, Sa99, Ma99, P01\\
2356$-$309 &   &  &  &   &  &X &0.165  &0.065  &0.67   &5.78     &Co01, Be92, Fa94  \\
%
%
\hline
\multicolumn{11}{l}{\footnotesize{$^a$ tentative redshift, Perlman priv. 
comm. (see NED notes).}}
\end{tabular}
\caption{TeV BL Lac candidates. S:Slew survey sample; Ex: Exosat sample; 
R: RGB sample; Ra: RASS sample; 1J: 1 Jy sample; oth.: others catalogues, 
see Donato et al. 2001 and references therein. None of the above sources 
is in the EMSS sample. The references report the basic source of data for the SEDs,
in addition to those in the NED database and the respective sample papers (Sect. 4.1). 
All the MeV-GeV upper limits  are from Fichtel et al. 1994 (first EGRET catalogue).
Ah00: Aharonian et al. 2000;
Ba94: Bade et al. 1994;
Ba00: Bauer et al. 2000;
Be92: Bersanelli et al. 1992;
Be01: Beckman et al. 2001;
Bk99: Buckley 1999;
Br89: Brown et al. 1989;
Br95: Brinkmann et al. 1995;
Br97: Brinkmann et al. 1997;
BS94: Brinkmann \& Siebert 1994;
C97 : Comastri et al. 1997;
Ca97: Catanese et a. 1997a; 
Ch99: Chadwick et al. 1999;
Ci95: Ciliegi et al. 1995;
Co01: Costamante et al. 2001;
Di96: Dingus et al. 1996;
E94:  Edelson et al. 1994;
F96:  Fruscione 1996;
Fa90: Falomo et al. 1990;
Fa94: Falomo et al. 1994;
Fo98: Fossati et al. 1998;
GAM95: Giommi, Ansari \& Micol 1995;
Ge94: Gear et al. 1994;
GS95: Ghosh \& Soundararajaperumal 1995;
GT96: George \& Turner 1996;
IN88: Inpey \& Neugebauer 1988;
Ke95: Kerrick et al 1995;
Ki97: Kifune et al. 1997;
La96: Lamer et al. 1996;
Li94: Litchfield et al. 1994;
Ma88: Madejski et al. 1988;
Ma99: Madejski et al. 1999;
Ne94: Neumann et al. 1994;
Pi93: Pian et al. 1993;
Pir00: F. Piron PhD thesis, 2000;
P01:  Padovani et al. 2001;
Ro99: Roberts et al. 1999;
Ro01: Robson et al. 2001;
Sa94: Sambruna et al. 1994;
SG99: Stevens \& Gear 1999;
St94: Stevens et al. 1994;
Si91: Sitko \& Sitko 1991;
Ta01: Tagliaferri et al. 2001 and references therein;
Te98: Terasranta et al. 1998;
To96: Tornikoski et al. 1996;
Wo00: Wolter et al. 2000 and references therein;
WW90: Worrall \& Wilkes 1990.
}
\end{center}
\vspace*{-0.2cm}
\end{table*}

\subsection{Selection of TeV candidate BL Lacs}

Fig. 4 shows the X--ray flux as a function of the radio flux 
for the BL Lacs objects in our ensemble.
Note the locations of the already TeV--detected sources:
they are among the brightest sources in both bands. 
This is not so obvious as it may seem at first sight, since a large X--ray 
to radio flux (hence, a lower radio emission for a given X--ray flux) 
indicates a large synchrotron peak frequency (see e.g. Fig. 8 in F98), 
which is the first requirement to emit in the TeV band (we must have 
many electrons energetic enough to emit copiously at TeV energies).
Consider also that the radio emission (at 5 GHz) must be produced
in a large region of the jet (not to be self--absorbed), much larger
than the part of the jet emitting high frequency radiation 
(as required by the very rapid variability).
Therefore the link between the radio/X--ray emission and the TeV flux
is more subtle.
 
We have interpreted this property in the following way:
to produce a large TeV flux by the IC process
we need many electrons of random Lorentz factors $\gamma\sim 10^5$--$10^6$.
These electrons emit synchrotron photons of energies 
$h\nu = 1.5 B(\gamma/10^5)^2(\delta/10)$ keV, with $B$ in Gauss.
The seed photons most effective to interact with these
electrons to produce TeV photons by the IC process
are in the IR--optical band, since photons of higher frequencies
scatter in the Klein Nishina regime.
We therefore propose that the radio flux  measures the level of
the relevant seed photons.
If this is the case, then, for a given X--ray flux, sources that 
are brighter in the radio band are more likely to be TeV emitters.
And indeed the objects already detected in the TeV band are 
bright both in the radio and in the X--ray bands.
We think this is exactly the effect of the ``trade--off" between high \gpeak
and the energy density of seed photons (see Sect. 3.2), since the objects with
higher $\nu_{\rm peak}$ are, on average, those with \emph{fainter} radio emission
for a given X--ray flux, i.e. those with lower \arx 
(see the correlation between \arx and $\nu_{\rm peak}$ in F98, 
Wolter et al. 1998, Costamante et al. 2001).
%

As shown in Fig. \ref{fxfr}, in the region  of high radio and X--ray flux
around the detected TeV objects there are also other sources, 
that we therefore consider  good candidates for TeV emission.
The extension of such region is, to some extent, subjective: 
the ``rectangle" in Fig \ref{fxfr} was drawn in order to 
include the already TeV detected sources and sources like 
Mkn 501 and Mkn 421 if they were at a redshift $\sim$ 0.1. 
In Fig. \ref{fxfr} an increase of the redshift corresponds to 
decrease the fluxes along the lines of constant $\alpha_{\rm RX}$ 
(the changing K--correction causes a negligible deviation, on this scale).
%

Fig. \ref{fxfo} shows how the BL Lac objects are placed
in the optical -- X--ray flux plane.
As the seed photons most effective for the TeV emission 
are in the IR--optical band, the optical flux could be in principle a better 
indicator of the density of seed photons than the radio flux.
However, due to the possible contamination (either as emission and 
absorption) from the host galaxy (and uncertainties on the intervening 
reddening medium), we consider the radio flux  a more reliable indicator 
of the low--energy non--thermal nuclear emission from these objects. 
Note that, in any case, all but one sources within the radio--X-ray 
rectangle are also within the optical--X--ray one. 
For a given X--ray flux, then,  a relatively high emission in both the radio 
and optical bands should represent a reliable indication of a large energy 
density of seed photons.  

Table 2 reports the list of the objects which are inside the two ``rectangles"
in Fig. \ref{fxfr}  and Fig. \ref{fxfo}, according to the values reported
in the respective samples (see Section 4.1), and  that therefore we consider the best 
candidates for a possible TeV detection.

\begin{table*}
\begin{center}
\begin{tabular}{|lllllllllllll|}
\hline
Name       &$L^\prime$   &$R$  &$B$ &$\Gamma$ &$\theta$ &$n$ &$\gamma_1$ 
&$\gamma_{\rm peak}$ &$\gamma_2$
& $F^b_{\rm (>40~GeV)}$  &$F^b_{\rm (>0.3~TeV)}$ &$F^b_{\rm (>1~TeV)}$ \\
      & erg s$^{-1}$ &cm   & G  &         &         &    &    &&&           &      & \\
\hline
0033$+$595      &1.0e41  &9.0e15 &0.8 &14   &3.0  &3.01 &1000 &5.3e4 &6.0e5  &15.0 / 2.93  &2.04 / 0.25 &0.48 / 0.04\\
0120$+$340      &4.4e41  &1.0e16 &0.6 &15   &3.0  &3.2  &700  &7.0e4 &2.5e5  &3.17 / 4.90  &0.28 / 0.30 &0.06 / --- \\
0136$+$391$^a$  &3.0e42  &1.0e16 &2.0 &15   &4.8  &3.6  &500  &4.5e3 &5.0e5  &5.22 / 4.00  &0.56 / 0.14 &0.12 / 2.7e-3\\
0214$+$517      &5.0e40  &8.0e15 &0.4 &13   &5.0  &3.95 &1500 &1.7e5 &6.0e5  &43.4 / 1.59  &5.93 / 0.07 &1.43 / 6.2e-3\\
0219$+$428      &5.0e42  &2.0e16 &1.4 &15   &3.0  &3.4  &700  &4.7e3 &8.0e4  &7.01 / 9.62  &0.14 / ---  &0.01 / --- \\
0229$+$200      &1.5e41  &1.0e16 &0.3 &15   &3.5  &3.5  &700  &2.8e5 &4.0e5  &7.67 / 3.81  &0.96 / 0.31 &0.21 / 4.0e-3\\          
0323$+$022      &1.0e41  &1.0e16 &0.9 &11   &3.7  &3.5  &800  &2.7e4 &1.8e5  &6.65 / 1.18  &0.84 / 0.01 &0.18 / ---   \\       
0414$+$009      &8.0e41  &1.0e16 &1.5 &14   &3.5  &3.2  &500  &9.8e3 &1.6e5  &2.91 / 3.42  &0.23 / 0.07 &0.04 / --- \\          
0548$-$322      &7.5e40  &8.0e15 &0.8 &12   &4.0  &3.1  &700  &5.0e4 &5.0e5  &31.9 / 1.56  &4.14 / 0.10 &0.91 / 0.015\\          
0556$-$383      &7.0e40  &8.0e15 &0.4 &16   &5.0  &3.2  &1500 &2.5e5 &2.5e5  &37.8 / 5.51  &5.84 / 0.42 &1.56 / ---    \\    
0647$+$250$^a$  &1.0e42  &1.2e16 &1.4 &16   &3.3  &3.6  &500  &1.0e4 &2.0e5  &6.16 / 8.74  &0.59 / 0.24 &0.12 / ---\\ 
0806$+$524      &1.0e42  &1.2e16 &1.5 &15   &4.0  &3.4  &300  &8.5e3 &7.0e4  &14.7 / 10.7  &1.36 / ---  &0.27 / --- \\          
0809$+$024$^a$  &8.0e41  &1.0e16 &1.3 &14   &4.8  &3.3  &1000 &1.2e4 &2.0e5  &6.08 / 2.20  &0.58 / 0.04 &0.12 / ---\\ 
0851$+$202      &5.0e42  &1.3e16 &5.0 &13   &3.5  &3.4  &150  &6.5e2 &7.e3   &23.7 / ---   &0.42 / ---  &0.03 / --- \\ 
1028$+$511      &2.7e41  &1.0e16 &1.0 &14   &3.0  &3.2  &800  &2.6e4 &2.0e5  &7.15 / 3.89  &0.43 / ---  &0.06 / ---\\
1011$+$496      &7.0e41  &2.0e16 &0.7 &12   &4.0  &3.4  &300  &1.9e4 &1.0e5  &1.67 / 3.31  &0.12 / 0.14 &0.02 / --- \\          
1101$-$232      &5.3e41  &7.0e15 &0.9 &16   &2.8  &3.1  &300  &5.3e4 &1.0e6  &6.67 / 10.2  &0.67 / 0.93 &0.15 / 0.18\\
1114$+$202      &4.5e41  &8.0e15 &1.5 &16   &3.0  &4.6  &6000 &6.0e3 &2.5e5  &10.1 / 8.51  &1.17 / 0.10 &0.28 / --- \\ 
1133$+$704      &2.0e41  &2.0e16 &0.8 &10   &5.0  &3.7  &400  &1.6e4 &1.7e5  &62.8 / 2.15  &8.50 / 0.03 &1.93 / --- \\          
1136$+$673      &4.0e41  &1.0e16 &1.0 &15   &4.2  &3.5  &1000 &2.4e4 &1.5e5  &7.30 / 5.40  &0.92 / 0.10 &0.21 / --- \\  
1215$+$303      &1.0e41  &1.0e16 &2.5 &11   &4.6  &3.8  &300  &3.4e3 &4.0e4  &4.06 / 0.07  &0.16 / ---  &0.02 / --- \\          
1218$+$304      &2.0e41  &1.0e16 &1.5 &16   &2.9  &3.9  &600  &1.4e4 &3.0e5  &6.36 / 5.82  &0.67 / 0.16 &0.15 / --- \\          
1417$+$257      &1.7e42  &1.0e16 &1.5 &15   &4.2  &3.4  &400  &8.4e3 &2.5e5  &3.76 / 6.88  &0.38 / 0.21 &0.08 / --- \\ 
1440$+$122      &2.6e41  &1.2e16 &0.7 &15   &4.2  &3.5  &500  &4.6e4 &5.0e5  &6.11 / 1.89  &0.78 / 0.09  &0.20 / 0.01\\     
1544$+$820$^a$  &6.0e41  &8.0e15 &1.2 &14   &4.0  &3.3  &400  &1.8e4 &4.0e5  &4.89 / 5.37  &0.54 / 0.22  &0.12 / ---\\ 
1553$+$113      &2.5e42  &3.0e16 &0.7 &15   &2.5  &3.6  &300  &1.3e4 &1.0e5  &8.92 / 22.3  &0.20 / 0.42  &0.02 / ---\\          
1722$+$119      &4.0e40  &1.5e16 &0.7 &10   &6.5  &3.6  &600  &2.6e4 &5.0e5  &76.6 / 1.06  &12.8 / 0.015 &3.52 / 1.0e-3 \\          
1727$+$502      &1.1e41  &1.0e16 &0.8 &10   &5.0  &3.5  &600  &3.0e4 &1.9e5  &38.7 / 2.64  &5.19 / 0.07  &1.23 / ---\\          
1741$+$196      &1.0e41  &1.0e16 &0.4 &13   &4.0  &3.8  &400  &1.3e5 &6.0e5  &31.6 / 4.31  &3.59 / 0.29  &0.84 / 0.01\\          
1959$+$650      &8.0e40  &1.0e16 &1.2 &13   &4.0  &3.6  &500  &1.9e4 &1.5e5  &56.7 / 2.08  &7.46 / 0.03  &1.74 / --- \\
2005$-$489      &3.5e41  &1.0e16 &1.2 &16   &3.0  &3.3  &600  &2.1e4 &5.0e5  &67.1 / 62.5  &5.14 / 2.67  &0.90 / 0.17\\
2200$+$428      &1.3e42  &5.0e15 &1.2 &14   &3.3  &4.3  &1000 &1.0e3 &2.0e5  &67.7 / 42.8  &3.32 / 0.17  &0.43 / ---\\
2356$-$309      &2.0e41  &7.0e15 &1.2 &16   &3.0  &3.01 &300  &6.0e4 &7.0e5  &7.64 / 3.30  &0.84 / 0.19  &0.12 / 0.03\\
%
%
\hline 
\end{tabular}
\caption{Input parameters for the SSC model, resulting values of 
$\gamma_{\rm peak}$, and predicted fluxes at high energies, above 40 GeV, 
0.3 TeV and 1 TeV, according to the parameterization of the SED adapted
from Fossati et al. (1998) (first number) and according to the SSC model 
discussed in Section 5.1 (second number).
$L^\prime$ is the intrinsic power (i.e. measured in the comoving frame),
$R$ the cross sectional radius of the emitting region,
$\Gamma$ is the bulk Lorentz factor,
$\theta$ is the viewing angle,
$n$ the slope of the particle distribution above the cooling
energy $\gamma_{\rm c} $ (see text),   
$\gamma_1$ and $\gamma_2$ are the extreme Lorentz factors of the 
injected particle distribution and 
$\gamma_{\rm peak}$ is the particle Lorentz factor of the electrons emitting 
most of the radiation (i.e. at the synchrotron and self--Compton peaks).
$^a$: $z=0.2$ assumed. 
$^b$ Fluxes in units of $10^{-11}$ photons cm$^{-2}$ s$^{-1}$.
} 
\end{center}
\end{table*}

\section{Prediction of the TeV flux}

To better quantify the predicted high energy flux of our
best candidates we have collected from the literature other data 
for all the objects listed in Table 2, in order to construct their SED.
We have then used two different methods to estimate their high energy emission.
First, for each source, we have applied a SSC model, as explained
below, aimed to fit the synchrotron component of their SED and to
predict the inverse Compton spectrum.
Then we have also calculated the predicted spectrum according to 
a slightly modified version of the parameterization given by F98, 
thought to describe the average SED of blazars.

Note that the predicted SEDs shown in Fig. 6 and the TeV fluxes
listed in Table  3 do not take into account the absorption of TeV 
photons by the IRB, since this is indeed one of the important
unknowns we would like to measure.

\subsection{Homogeneous Synchrotron Self--Compton model}

We have applied a homogeneous, one--zone synchrotron self--Compton
model to our best TeV candidate BL Lacs.
This model is ``one--zone version" of the model in Spada et al. (2001) 
and is described in detail in Ghisellini, Celotti \& Costamante (2001).
The main characterizing feature of this model is the assumption
of an particle injection mechanism lasting for a finite time.
In this case the emitting particle distribution never reaches a
complete steady state.
A physical scenario where such a behavior naturally occurs is that of
internal shocks (see e.g. Piran 1999; Ghisellini 1999; Spada et al. 2001), 
i.e. collisions of different parts of the jet plasma moving at slightly 
different speeds, naturally leading to shocks lasting for a finite time.
This scenario is suggested by the rapid variability always
present in BL Lac objects in general and HBL in particular,
especially at high energies.
We briefly outline here the other main assumptions of the model.

We assume that the jet is conical, with half opening angle $\psi\simeq 1/\Gamma$,
and approximate the emitting region  as a cylinder, of radius $R$ ($=\psi z$, where 
$z$ is the distance along the jet axis) and width
$\Delta R^\prime=R/\Gamma$ (in the comoving frame, here $\Gamma$
is the bulk Lorentz factor). This corresponds to assume 
$\Delta R^\prime$ constant
up to the first collision, in the internal shocks scenario (Spada et al. 2001).  
We derive the particle distribution assuming that the
acceleration mechanism is in the form of a continuous
injection of relativistic particles distributed in energy
as a power law of index $n-1$ between $\gamma_1$ and $\gamma_2$.
The injection is assumed to last for a finite time, set equal
to $t_{\rm inj}=\Delta R^\prime/c$.
We define $\gamma_{\rm c}$ as the energy of those electrons that can cool
in the injection time $t_{\rm inj}$, i.e. $t_{\rm cool}(\gamma_{\rm c})=t_{\rm inj}$.
At energies greater than $\gamma_{\rm c}$, particles radiatively cool, and
the distribution reaches a steady state in a time smaller than $t_{\rm inj}$.
As a consequence, the emitting particle distribution is assumed
to be a power law of index $n$ above $\gamma_{\rm c}$.
Below this value, there can be different cases according if
$\gamma_{\rm c}$ is greater or smaller than $\gamma_1$.
If $\gamma_{\rm c}>\gamma_1$, we have $N(\gamma) \propto \gamma^{-(n-1)}$ 
between $\gamma_1$  and $\gamma_{\rm c}$.
Alternatively, if $\gamma_{\rm c} < \gamma_1$, 
then $N(\gamma) \propto \gamma^{-2}$ 
between $\gamma_{\rm c}$ and $\gamma_1$.
We further assume that, below the minimum between $\gamma_1$
and $\gamma_{\rm c}$, $N(\gamma)\propto \gamma^{-1}$.
According to these assumptions, the random Lorentz factor 
$\gamma_{\rm peak}$ of the electrons emitting most of the radiation 
(i.e. emitting at the peaks of the SEDs) is determined by  
the importance of radiative losses and can have values
within the range $\gamma_1$--$\gamma_2$.
Its value is listed in Table 3.
The source is assumed to emit an intrinsic luminosity $L^\prime$
and is assumed to be observed with the viewing angle $\theta$.
All these input parameters are listed in Table 3, together with the predicted high
energy photon fluxes above three representative frequencies.
The resulting fits are shown in Fig. 6 as solid lines.

In conventional SSC models (Tavecchio, Maraschi \& Ghisellini 1998), 
the knowledge of the location of the two peaks and their fluxes, together 
with the variability timescale, suffices to derive unambiguously all 
relevant physical parameters.
Here, instead, we lack two important observables, namely the
frequency and flux of the Compton peak.
We therefore need to supply two additional relations with respect
to the ``standard" SSC model.

One of these relations comes from our assumption of finite injection
time, resulting in a relation between $\nu_{\rm peak}$ and
the magnetic field.
In fact, if synchrotron losses are relevant (as in the sources already 
detected at TeV energies, for which the high energy component never exceeds
the synchrotron one), we have a constraint on the value of 
the magnetic field, controlling the radiative losses timescales and
hence the value of $\gamma_{\rm peak}$.

For the second unknown to be provided, we decided to limit the value
of the beaming Doppler factor $\delta$ within the relatively narrow range
$9<\delta<20$.

Note, however, that also within the previous assumptions, the $\gamma$--ray
flux predictions still have a large uncertainty, mainly due to 
the dependence of the flux from the emission volume 
(i.e. from $R$ and $\Gamma$). In our model, however, these 
are not completely free parameters, since on one hand they must
be compatible with the variability timescales, and on the other they control
the injection timescale $t_{\rm inj}$. 
The dependence of the inverse Compton flux from $R$ and $\Gamma$
is therefore complex. For instance, for a given $L^{\prime}$ and $B$,
an increase of $R$ makes the Compton flux to decrease ($\propto R^{-2}$),
but it makes also $t_{\rm inj}$ to increase, leading to a smaller value of 
$\gamma_{\rm c}$ ($=\gamma_{\rm peak}$), and then to a smaller synchrotron peak frequency.
To compensate for that, it would be necessary to decrease $B$ 
(to increase the cooling time), but this leads to an {\it increase} of the
Compton flux.

Our choice has been to use values of $R$ around $\sim10^{16}$ cm,
which are the typical values obtained for sources 
with good monitoring and  with SEDs well sampled
also at high energies, like Mkn 421 and 1ES 2155--304.


Another point of uncertainty is the exact determination 
of $\nu_{\rm peak}$ and the corresponding synchrotron peak flux.
For many sources this is provided by the {\it Beppo}SAX observations,
which can be fitted by a broken power law, immediately yielding
$\nu_{\rm peak}$.
For other sources of poorly known SED the determination of 
$\nu_{\rm peak}$ is more uncertain, and for these sources
the predicted SEDs of course suffer from this uncertainty
(see e.g. the SED of 0214+517, 1440+122 and 1544+820, for which there
is no information of the slope of the X--ray spectra and only a few
observations in the IR--optical bands).

The applied model is aimed to reproduce the spectrum 
originating in a limited part of the jet, thought to
be responsible of most of the emission. 
This region is necessarily compact, since it must account
for the fast variability shown by all blazars, especially
at high frequencies.
Therefore the radio emission from this compact regions
is strongly self--absorbed, and the model cannot account
for the observed radio flux.
\setcounter{figure}{5}
\begin{figure*} 
\psfig{figure=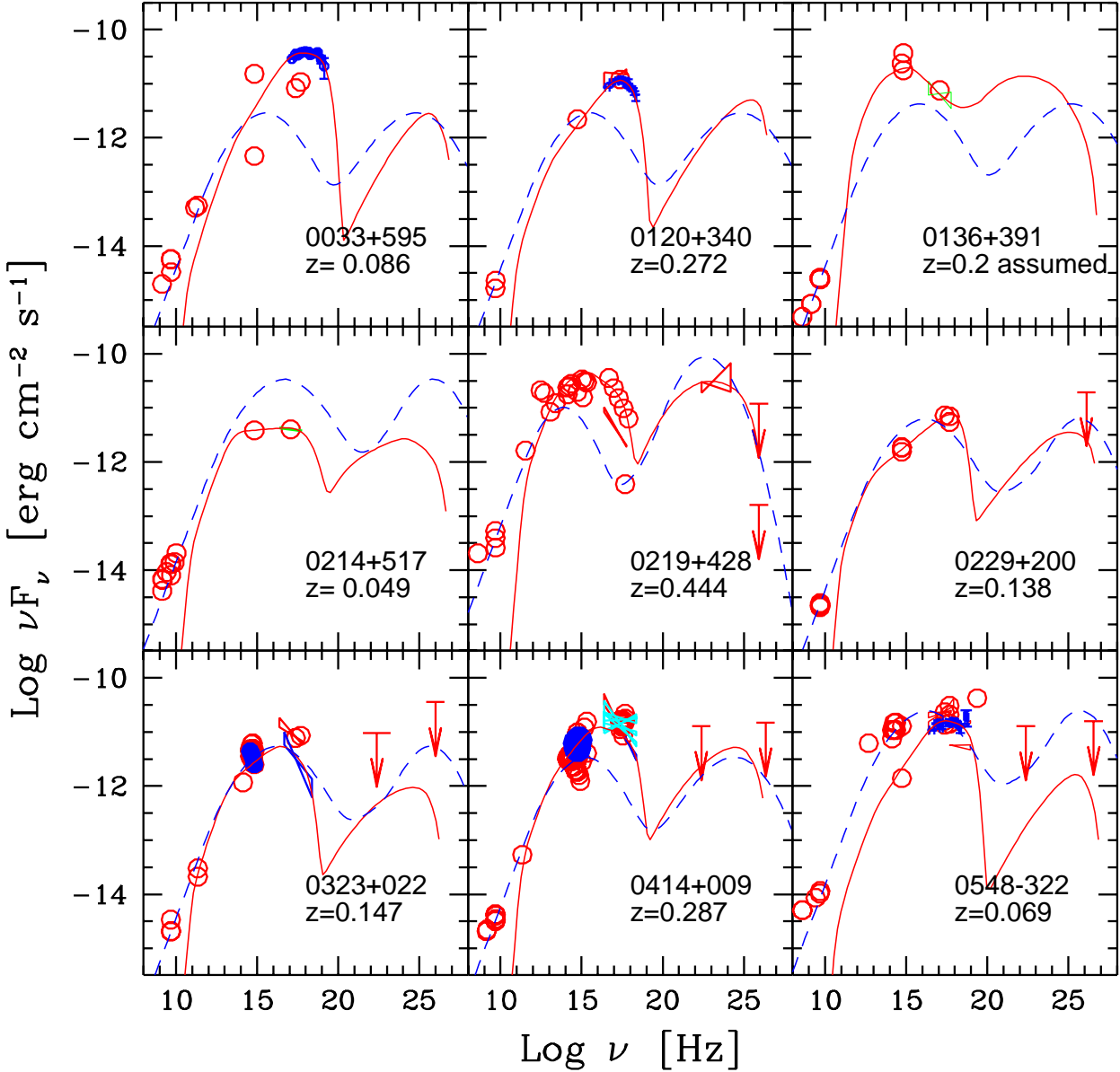,width=17.5cm,height=17.5cm}
\vskip -1 true cm
\caption{a):
SEDs of our best candidates for TeV emission.
The solid lines refer to the SSC model as explained in Section 5.1.
Dashed lines correspond to the phenomenological prescription of
Fossati et al. (1998), as slightly modified by Donato et al. (2001)
and in this paper.
Sources of data listed in Table 2.
}
\end{figure*} 

\setcounter{figure}{5}
\begin{figure*} 
\psfig{figure=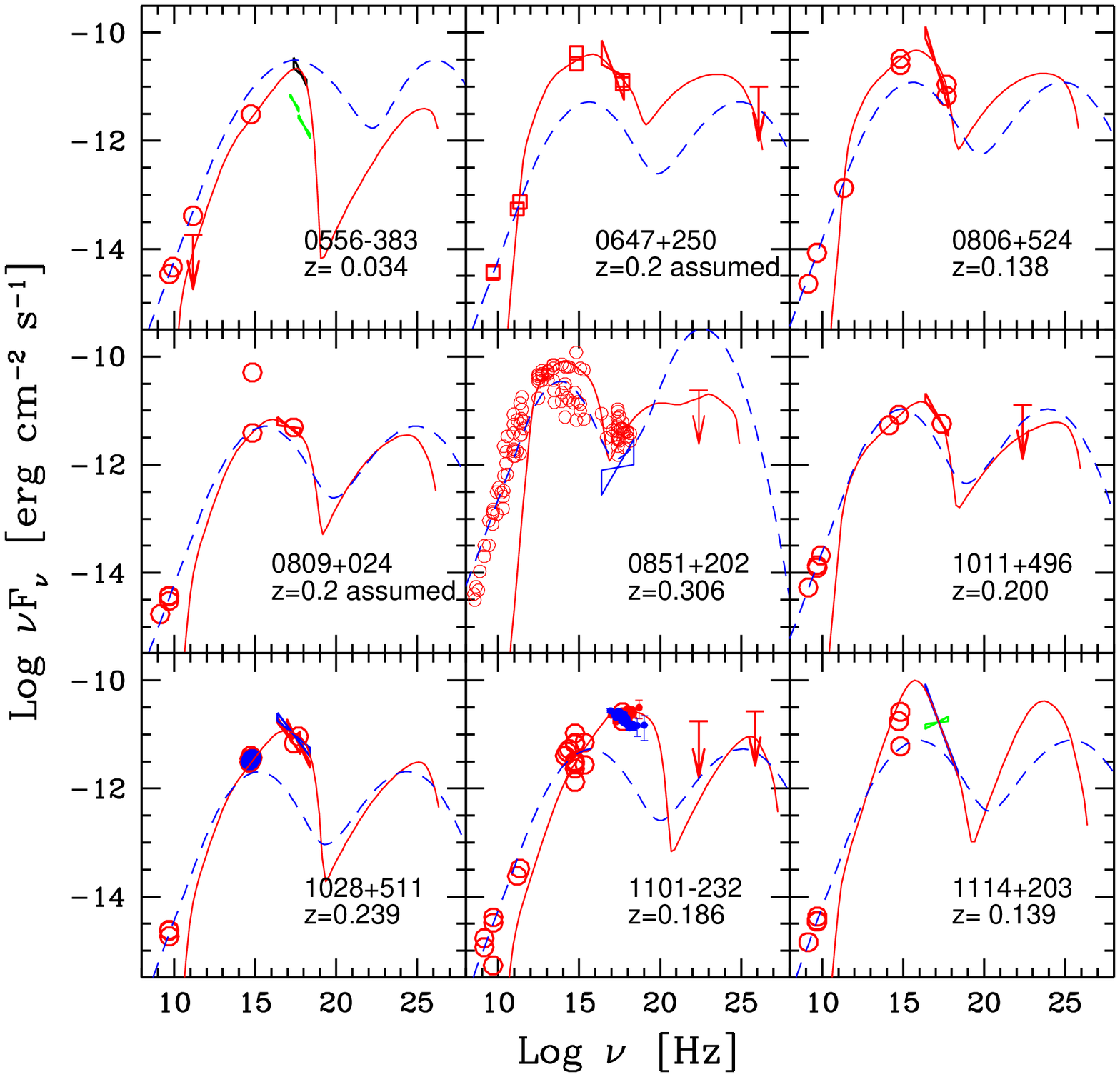,width=17.5cm,height=17.5cm}
\vskip -1 true cm
\caption{b):
SEDs of our best candidates for TeV emission.
The solid lines refer to the SSC model as explained in Section 5.1.
Dashed lines correspond to the phenomenological prescription of
Fossati et al. (1998), as slightly modified by Donato et al. (2001)
and in this paper.
Sources of data listed in Table 2.
}
\end{figure*} 

\setcounter{figure}{5}
\begin{figure*} 
\psfig{figure=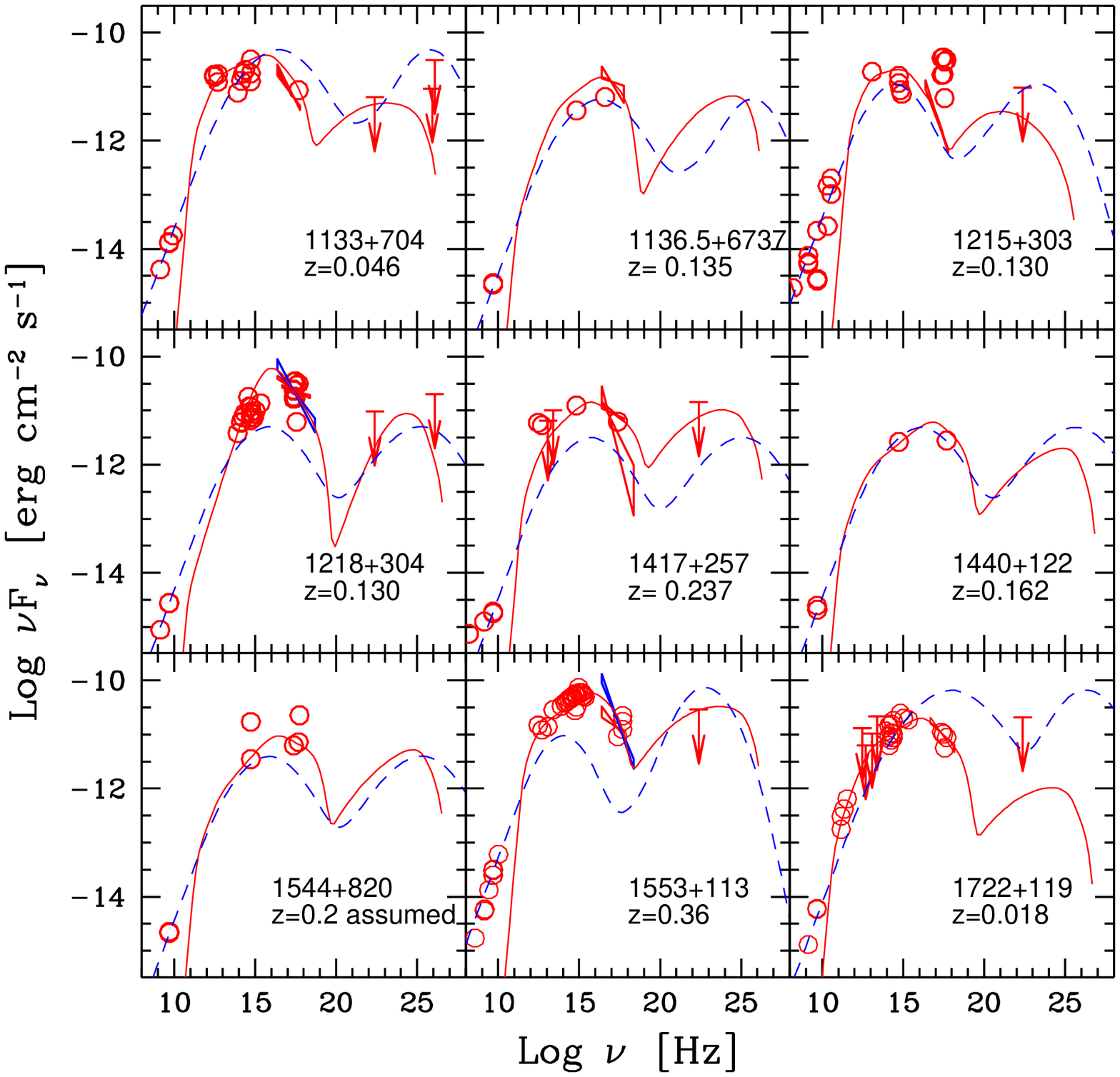,width=17.5cm,height=17.5cm}
\vskip -1 true cm
\caption{c):
SEDs of our best candidates for TeV emission.
The solid lines refer to the SSC model as explained in Section 5.1.
Dashed lines correspond to the phenomenological prescription of
Fossati et al. (1998), as slightly modified by Donato et al. (2001)
and in this paper.
Sources of data listed in Table 2.
}
\end{figure*} 

\setcounter{figure}{5}
\begin{figure*} 
\psfig{figure=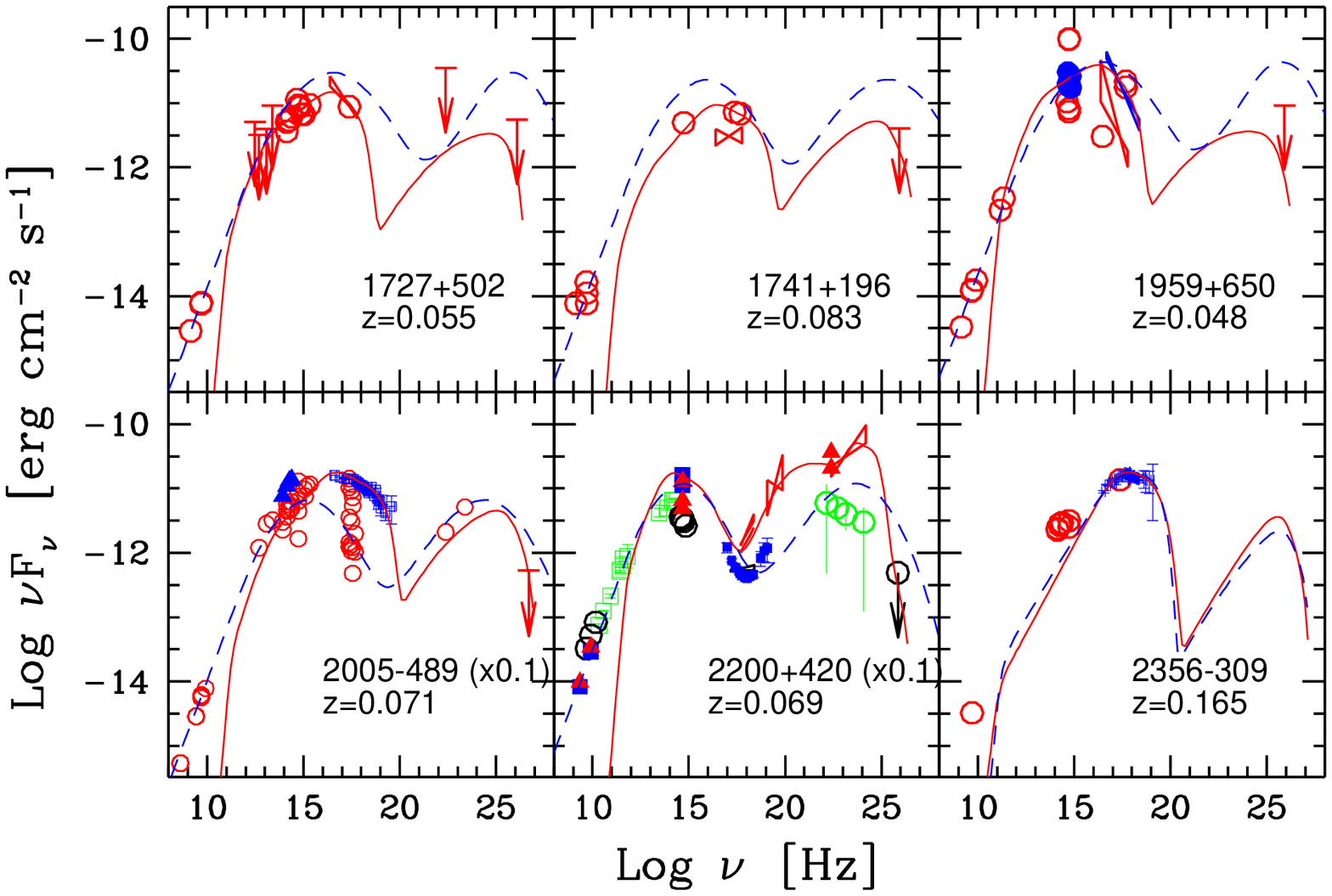,width=17.5cm,height=17.5cm}
\vskip -6 true cm
\caption{d):
SEDs of our best candidates for TeV emission.
The solid lines refer to the SSC model as explained in Section 5.1.
Dashed lines correspond to the phenomenological prescription of
Fossati et al. (1998), as slightly modified by Donato et al. (2001)
and in this paper.
Sources of data listed in Table 2.
}
\end{figure*}

\subsection{Fossati et al. (1998) description of the SED}

Fossati et al. (1998) proposed a simple phenomenological 
description of the average SED of blazars based on their 
bolometric observed luminosity, thought to be well traced by the radio
luminosity. In their parameterization, the radio luminosity is assumed to
be the key parameter, determining the peak frequency of the synchrotron 
spectrum and the relative importance of the inverse Compton power.
More recently, Donato et al. (2001) have revisited this
parameterization, assuming a slightly different relation between 
the radio power, the synchrotron peak frequency and the inverse 
Compton luminosity, but only for objects below a radio luminosity of 
$10^{43}$ erg s$^{-1}$.
With this new parameterization, objects of low power are
assumed to have equal luminosities in the synchrotron and in the 
self--Compton components of their spectra, and the ratio of the peak 
frequencies ($\nu_{\rm c}/\nu_{\rm s}$) was allowed to increase for increasing $\nu_{\rm s}$
(as expected in a SSC scenario).

To improve this parameterization also for objects with even 
lower radio luminosity (i.e. higher $\nu_{\rm peak}$), 
we have further modified the Donato et al. prescription
for sources with radio luminosity below $10^{41.2}$ erg s$^{-1}$,
to mimic some effects of the Klein--Nishina regime.
The first modification involves the ratio $\nu_{\rm c}/\nu_{\rm s}$,
since in the Klein--Nishina regime, $\nu_{\rm s}\propto \gamma^2_{\rm peak}$
while $\nu_{\rm c}\propto \gamma_{\rm peak}$.
Then the ratio 
$\nu_{\rm c}/\nu_{\rm s}\propto \gamma^{-1}_{\rm peak}\propto \nu_{\rm s}^{-1/2}$.
Below $L_{\rm R}=10^{41.2}$ erg s$^{-1}$, we then decrease $\nu_{\rm c}/\nu_{\rm s}$
assuming that  
$\log(\nu_{\rm c}/\nu_{\rm s})=9.4-0.8\times [41.2-\log L_{\rm R}]$.
%
The other modification concerns the width of the parabola representing 
the Compton peak,
which is reduced with respect to the synchrotron one
($\sigma^2_{\rm Comp}=\sigma^2_{\rm sync}/2$;  see Fossati et al. (1997), appendix A, 
and F98 for a description of the parameters). 
In this way the resulting SED shapes provide a better representation of 
the individual SEDs for the highest $\nu_{\rm peak}$ sources. 

We have applied this parameterization to all the objects in our sample, 
and the resulting fits are shown as dashed lines in 
Fig. 6. From these we obtained the photon fluxes listed in Table  3, 
integrating above the corresponding threshold energy.

From the comparison with the homogeneous SSC models, we can 
see that the F98 parameterization tends to overestimate
the Compton $\gamma$--ray emission, although it agrees with the
existing upper limits in the GeV and TeV bands in all cases
but 0851+202, 1133+704 and 1553+113. 
This is not unexpected, since, by construction, the Compton power
is never less than the synchrotron one and the assumed parabolic shape
does not account for sharp high energy cut-offs.
This parameterization however, which is built to describe the average SED of 
sources of equal synchrotron and self--Compton power, 
is in good agreement with the flux measurements of the already detected TeV
sources (see Fig. \ref{tevdet}).

\section{Discussion and conclusions}

With new Cherenkov telescopes foreseen to operate in the next few 
years the TeV extragalactic astronomy is entering its adulthood.
A tenfold increase in sensitivity (expected for the forthcoming installations)
would mean the possible detection of $\sim$ 100 BL Lacs, if the
counts of BL Lac object at TeV energies are roughly Euclidean
(in the bright flux end) and neglecting absorption by IRB.
Because of IRB absorption, the counts will be flatter than Euclidean, but
the lower energy threshold of some new instruments may compensate   
for the extragalactic absorption,
as well as favoring the detection of
slightly less blue objects. Therefore  many more sources are expected to be 
detectable by the new telescopes, and this
motivated us to study which kind of BL Lac objects is more
likely to be detected at high energies.  

Our findings can be summarized very simply once we realize that,
to produce a strong TeV emission, the inverse Compton process needs
a sufficient number of both very high energy electrons and soft seed photons.
Therefore we require both a strong X--ray flux and a sufficiently
strong radio--through--optical flux.
Since the optical flux can be contaminated, especially in low redshift
sources, by the underlying host galaxy, our  sources
are primarily selected as bright both in the X--ray and radio bands.
All but one of these sources are also bright in the optical.

With respect to the previous work by Stecker et al. (1996),
our criterium introduces the further requirement that the
source must be a relatively strong radio emitter.
The other difference is that we considered not only the
Einstein Slew survey sample of BL Lacs, but several other
BL Lac samples.

Besides selecting the best candidates through their location 
in the radio -- X-ray
flux plane (a criterium  largely model independent),
we have also tried to quantify the level of the expected high 
energy emission for each selected source, by applying a one--zone
synchrotron self--Compton model and also the phenomenological
description of the SED of Fossati et al. (1998), slightly
modified to better account of the average SED of low power BL Lacs.
The latter model, by construction, assumes equal power between 
the synchrotron and the inverse Compton components of the SEDs,
and almost always predicts larger high energy fluxes than the
SSC model. 
In the SSC model, in fact, the Compton dominance is not fixed a priori,
but found by fitting the synchrotron part of the spectrum,
which fixes the value of the magnetic field.
We stress that the F98 prescription was
designed including also the SEDs of those BL Lacs already
detected in the TeV band (and indeed is in good agreement with their
TeV flux levels), and therefore seems  more appropriate 
to predict the TeV flux of sources in high state.
The adopted SSC model, instead, is designed to fit the 
known synchrotron part of the SED, which is often representative
of a more ``normal" or quiescent state.
This explains the sometimes large discrepancy between the predicted
fluxes of the two models.
Since BL Lac objects are among the most variable sources, especially
at high energies, the two foreseen flux levels could be thought of
as an approximate range of variability, and
the average flux could be considered as a measure of the
probability to find the source in a particular TeV state.

We would like to stress, anyway, that the uncertainties on the key 
parameters we used for the model and the non--simultaneity of the fitted data
lead to large uncertainties in the predicted TeV flux, sometimes of the same 
order of the differences between the two adopted models.
Our predicted fluxes, therefore, also in the case of the SSC model,
must be considered as ``best guesses" on the high energy emission 
from these objects.
Within these limits, the SSC model provides more information
than the phenomenological parameterization, since it gives also
the expected shape of the high energy spectrum.  

Since the level of the synchrotron X--ray flux measures, in our scenario 
(as well as in the Stecker et al. 1996 one), the number of
TeV energy electrons, the X--ray monitoring of our candidates
is particularly useful to catch sources in high TeV states, as
already partially done through the All Sky Monitor (ASM) onboard
the {\it Rossi}XTE satellite.

Besides the fluxes above 300 GeV, the most common threshold of 
present Cherenkov
telescopes, we  also give our estimates above 40 GeV, which is 
approximately the energy threshold of CELESTE and of  forthcoming 
observatories like HESS and MAGIC  (with VERITAS at $\sim50$ GeV, 
Weekes 1999).
Emission at these energies is much less absorbed by the 
cosmic infrared background, giving the opportunity to 
see more distant sources and study their intrinsic spectrum in
an unabsorbed band.
In addition this energy range will link ground based Cherenkov
observations and the data coming from satellites, such as 
AGILE and GLAST, observing from a few tens of MeV to a few tens of GeV.

As a final note, we warn that our flux estimates \emph{do not}
include the possible absorption due to the infrared background,
since we preferred to be independent of this factor.
In fact, we focussed on the conditions for the TeV emission, providing a list
of possible sources, in order to allow an independent test of the
IR absorption effects. 
This differs with respect to the flux estimates in Stecker et al. (1996), 
which instead account for the IR background absorption.
However, since most of the photon flux is expected to be emitted 
at energies below 1 TeV
(as can be seen in Table 3, comparing the values above 0.3 and 1 TeV),
the reported fluxes should not be much affected for sources up to 
$z\sim 0.1$,
according to the present estimates on the IR background
(see Stecker 2001 and references therein).

\begin{acknowledgements}
We thank the referee, H. Krawczynski, for his constructive suggestions.
We thank D.A. Smith, A. Djannati--Atai, I. de la Calle Perez, F. Piron 
and A. Celotti for useful discussions.
This research has made use of the NASA/IPAC Extragalactic Database (NED)
which is operated by the Jet Propulsion Laboratory, Caltech, under contract
with the National Aeronautics and Space Administration.
\end{acknowledgements}


\begin{thebibliography}{}
\bibitem[]{} Aharonian F.A., Akhperjanian A.G., Barrio J.A., et al., 1997, A\&A, 327, L5
\bibitem[]{} Aharonian F.A., Akhperjanian A.G., Barrio J.A., et al., 2000, A\&A, 353, 847  
\bibitem[]{} Aharonian F.A., 2000,  New Astron. 5, p. 377-395, astro--ph/0003159
\bibitem[]{} Aharonian F.A., Timokhin A.N., Plyasheshnikov A.V., 2001, 
        astro--ph/0108419 
\bibitem[]{} Aharonian F.A., 2001b, 27th International Cosmic ray Conference, Hamburg, Germany, in press
\bibitem[]{} Amelino-Camelia G. \& Piran T., 2001, Phys.Rev., D64, 036005, astro--ph/0008107
\bibitem[]{} Bade N., Fink H. \&  Engels D.,  1994, A\&A, 286, 381
\bibitem[]{} Bade N., Beckmann V., Douglas N.G., et al., 1998, A\&A, 334, 459
\bibitem[]{} Bauer F.E., Condon J.J., Thuan T.X.,  Broderick J.J., 2000, ApJS, 129, 547
\bibitem[]{} Beckmann V., Wolter A., Celotti A., et al., 2001, A\&A, in press
\bibitem[]{} Berezinsky V., 2001,  9th Int. Workshop ``Neutrino Telescopes" 
             (astro--ph/0107306)
\bibitem[]{} Bersanelli M., Bouchet P., Falomo R., Tanzi E.G., 1992, AJ, 104, 28
\bibitem[]{} Brinkmann W. \& Siebert J., 1994, A\&A, 285, 812
\bibitem[]{} Brinkmann W., Siebert J., Reich W., et al. 1995, A\&AS, 109, 147
\bibitem[]{} Brinkmann W., Yuan W., Siebert J., 1997, A\&A, 319, 413
\bibitem[]{} Brown L.M., Robson E.I., Gear W.K., et al., 1989, ApJ 340, 129
\bibitem[]{} Buckley J.H., 1999, APh, 11, 119
\bibitem[]{} Burstein D. \& Heiles C., 1982, AJ,  87, 1165 
\bibitem[]{} Catanese M., Akerlof C. W., Biller S. D., et al., 1997a, ApJ, 480, 562
\bibitem[]{} Catanese M., Bradbury S.M., Breslin A.C., et al., 1997b, ApJL, 487, L147
\bibitem[]{} Catanese M. \& Weekes T.C., 1999, PASP, 111, 1193
\bibitem[]{} Celotti A., 2001, {\it Blazar Physics and Demographics}, 
             eds. M.C. Urry, P. Padovani, ASP, 227, p. 105
\bibitem[]{} Chadwick P.M., Lyons K., MCComb T.J.L., et al., 1999, ApJ, 521, 547
\bibitem[]{} Ciliegi P., Bassani L., Caroli E., 1995, ApJ, 439, 80
\bibitem[]{} Comastri A., Fossati G., Ghisellini G., Molendi S., 1997, ApJ, 480, 534
\bibitem[]{} Costamante L., Ghisellini G., Tagliaferri G. et al., 2001a, 
             {\it Stellar Endpoints, AGN 
             and the Diffuse Background}, in press (astro--ph/0001410)
\bibitem[]{} Costamante L., Ghisellini G., Giommi P., et al., 2001b, A\&A, 371, 512
\bibitem[]{} Dermer C., Schlickeiser R. \& Mastichiadis A. 1992, A\&A, 256, L27 
\bibitem[]{} Dingus B.L., Bertsch D.L., Digel S.W., et al., 1996, ApJ, 467, 589
\bibitem[]{} Djannati--Ata\"{\i} A.,  Piron F., Barrau A., et al., 1999, A\&A, 350, 17
\bibitem[]{} Donato D., Ghisellini G., Tagliaferri G. \& Fossati G., 2001, A\&A, 375, 739
\bibitem[]{} Edelson R.A., 1994, AJ, 94, 1150
\bibitem[]{} Fabian A.C., Celotti A., Iwasawa K., McMahon R.G., Carilli C.L, 
             Brandt W.N., Ghisellini G., Hook, I.M., 2001a, MNRAS, 323, 373 
\bibitem[]{} Fabian A.C., Celotti A., Iwasawa K. \& Ghisellini G., 2001b, MNRAS, 324, 628  
\bibitem[]{} Falomo R. \& Treves A., 1990, PASP, 102, 1120
\bibitem[]{} Falomo R., Scarpa R., Bersanelli M.,  1994 ApJS, 93, 125
\bibitem[]{} Fichtel C.E., Bertsch D.L., Chiang J., et al., 1994, ApJS, 94, 551
\bibitem[]{} Fossati G., Celotti A., Ghisellini G., Maraschi L., 1997, MNRAS, 289, 136
\bibitem[]{} Fossati G., Maraschi L., Celotti A., Comastri A., Ghisellini G., 1998, MNRAS, 299, 433 (F98) 
\bibitem[]{} Fossati G., Celotti A., Chiaberge M., 2000, ApJ, 541, 153
\bibitem[]{} Fruscione A., 1996, ApJ, 459, 509
\bibitem[]{} Gaidos J.A., Akerlof C.W., Biller S.D., et al., 1996, Nature, 383, 319 
\bibitem[]{} Gear  W.K., Stevens J.A., Hughes D.H., et al.,  1994, MNRAS, 267, 167
\bibitem[]{} George I.M. \& Turner T.J., 1996, ApJ, 461, 198
\bibitem[]{} Ghisellini G. \& Madau P., 1996, MNRAS, 280, 67 
\bibitem[]{} Ghisellini G., Celotti A., Fossati G., Maraschi L., Comastri A., 1998, MNRAS, 301, 451 
\bibitem[]{} Ghisellini G., 1999, {\it The Active X--ray Sky: Results from 
        BeppoSAX and Rossi--XTE},  Nucl. Physics B Proc. Supp. Eds.: 
        L. Scarsi, H. Bradt, P. Giommi \& F. Fiore, p. 397 
\bibitem[]{} Ghisellini G., Celotti A. \& Costamante L., 2001,
             submitted to A\&A
\bibitem[]{} Ghosh K. \& Soundararajaperumal S., 1995, ApJS, 100, 37
\bibitem[]{} Giommi P., Ansari S.G. \&  Micol A., 1995, A\&AS, 109, 267  
\bibitem[]{} Giommi P., Barr P., Garilli B., Maccagni D., Pollock A.M.T., 
             1990, ApJ, 356, 432
\bibitem[]{} Giommi P., Padovani P. \& Perlman E., 1997, MNRAS, 317, 743
\bibitem[]{} Hartman R.C., Bertsch D.L., Bloom S.D., et al., 1999, ApJS,  123, 79
\bibitem[]{} Hauser G.H. \& Dwek E., 2001, ARAA, Vol. 39, (astro--ph/0105539)
\bibitem[]{} Horan D., VERITAS collaboration, 2000, HEAD meeting, No 32, 05.03
\bibitem[]{} Horan D., et al., 2001, AIP Conference Proceedings 578, Gamma-Ray Astrophysics
2001, ed. S.Ritz, N.Gehrels and C.R.Schrader, p. 324.
\bibitem[]{} Inpey C.D. \& Neugebauer G., 1988, AJ 95, 307;
\bibitem[]{} Kerrick A.D., Akerlof C.W., Biller S., et al., 1995, ApJ, 452, 588
\bibitem[]{} Kifune T.,  Dazeley S.A.,  Edwards P.G., et al., 1997, astro--ph/9707001
\bibitem[]{} Krawczynski H., Sambruna R., Kohnle A., et al., 2001, ApJ, 559, 187
\bibitem[]{} Krennrich F., Badran H.M., Bond I.H., et al., 2001, ApJL, 560, 45  
\bibitem[]{} Lamer G., Brunner H., Staubert R., 1996, A\&A, 311, 384
\bibitem[]{} Laurent--Muehleisen S.A., Kollgaard R.I., Feigelson E.D., 
             Brinkmann W., Siebert J., 1999, ApJ, 525, 127         
\bibitem[]{} Litchfield S.J., Robson E.I. \&  Stevens J.A., 1994, MNRAS, 270, 341
\bibitem[]{} Macomb D.J., Akerlof C.W., Aller H.D., et al., 1995, ApJL, 449, 99
\bibitem[]{} Madejski G.M. \& Schwartz D.A., 1988, ApJ 330, 776
\bibitem[]{} Madejski G.M., Sikora M., Jaffe T., et al.,  1999, ApJ, 521, 145
\bibitem[]{} Mannheim K., 1993, A\&A, 269, 67
\bibitem[]{} Maraschi L., Fossati G., Tavecchio F., et al., 1999, ApJL, 526, 81
\bibitem[]{} Muecke A. \& Protheroe R.J., 2000, astro-ph/0004052
\bibitem[]{} Neshpor Y.I. et al., 1998, Astron. Letts., 24, 134 
\bibitem[]{} Neumann M., Reich W., Fuerst E., et al., 1994, A\&AS, 106, 303
\bibitem[]{} Nishiyama T. et al., 1999, in Proc. of the 26th ICRC, ed. 
             D. Kieda et al., 3, 370 
\bibitem[]{} Padovani P.\& Giommi P., 1995, ApJ 444, 567   
\bibitem[]{} Padovani P., Costamante L., Giommi P., et al., 2001, MNRAS, in press
\bibitem[]{} Perlman E.S., Stocke J.T., Schacter J.F. et al., 1996, ApJS, 104, 251
\bibitem[]{} Pian E.\& Treves A., 1993, ApJ, 416, 130
\bibitem[]{} Pian E., Vacanti G., Tagliaferri G., et al., 1998, ApJL, 492, 17
\bibitem[]{} Piran, T., 1999, Phys. Rep., 314, 575
\bibitem[]{} Piron F., 2000, PhD Thesis, University of Paris--Sud (Paris XI)
\bibitem[]{} Protheroe R.J. et al., 1998,
       {\it 25th International Cosmic Ray Conference}. M.S. Potgieter, 
       B.C. Raubenheimer \& D.J. van der Walt Eds. (Singapore River Edge, NJ). 
       World Scientific, p. 317 (astro-ph/9710118)
\bibitem[]{} Protheroe R.J. \& Meyer H., 2000, Physics Letters B, 493, 1
\bibitem[]{} Punch M., Akerlof C.W., Cawley M.F., et al., 1992, Nature, 358, 477  
\bibitem[]{} Rachen J.P., 1999, GeV--TeV Gamma--Ray Astrophysics Workshop
             (Snowbird, USA), p. 41 
\bibitem[]{} Rector T., Stocke J.T., Perlman E.S., Morris S.L., Gioia I., 
             2000, AJ, 120, 1626
\bibitem[]{} Roberts M.D., McGee P.,  Dazeley S.A., 1999, astro--ph/9902008
\bibitem[]{} Robson E.I., Stevens J.A., Jenness T., 2001, MNRAS, in press, astro--ph/0107112
\bibitem[]{} Sambruna R.M., Barr P., Giommi P., Maraschi L., Tagliaferri G., 
             Treves A., 1994, ApJS, 95, 371
\bibitem[]{} Sambruna R.M., Ghisellini G., Hooper E., et. al., 1999, ApJ, 515, 140
\bibitem[]{} Sikora M., Begelman M.C. \& Rees M.J., 1994, ApJ, 421, 153
\bibitem[]{} Sitko M.L. \& Sitko A.K., 1991, PASP, 103, 160
\bibitem[]{} Spada M., Ghisellini G., Lazzati D. \& Celotti A., 
             2001, MNRAS, 325, 1559 
\bibitem[]{} Stecker F.W., De Jager O.C. \& Salamon M.H., 1996, ApJ, 473, L75
\bibitem[]{} Stecker F.W., 2001, Proc. IAU Symp. 204, {\it The Extragalactic
    Background and its Cosmological Implications}, M. Harwit and M.G. Hauser Eds., astro--ph/0010015
\bibitem[]{} Stevens J.A., Litcfield S.J., Robson E.I., et al., 1994, ApJ, 437, 91
\bibitem[]{} Stevens J.A. \& Gear W.K., 1999, MNRAS, 307, 403
\bibitem[]{} Stickel M., Fried J.W., Kuhr H., 1993, A\&AS, 98, 393
\bibitem[]{} Tagliaferri G., Ghisellini G., Giommi P., et al.,  2001, A\&A, 368, 38 
\bibitem[]{} Takahashi T., Kataoka J., Madejski G., et al., 2000, ApJL, 542, 105
\bibitem[]{} Tavecchio F., Maraschi L. \& Ghisellini G., 1998. ApJ, 509, 608 
\bibitem[]{} Teraesranta H., Tornikoski M., Mujunen A., et al., 1996, A\&AS, 116, 157
\bibitem[]{} Urry C.M., Sambruna R.M., Worrall D.M. et al., 1996, ApJ, 463, 424
\bibitem[]{} Weekes T., 2000, Proceedings oof GeV-TeV Gamma Ray Astrophysics Workshop 
(Snowbird), Eds Dingus et al.,Vol. 515, p.3  (astro-ph/9910394)
\bibitem[]{} Wolter A., Comastri A., Ghisellini G., et al., 1998 ,A\&A, 335, 899
\bibitem[]{} Wolter A., Tavecchio F., Caccianiga A., Ghisellini G., Tagliaferri G., 2000, A\&A, 357, 429
\bibitem[]{} Worrall D.M. \& Wilkes B.J., 1990, ApJ, 360, 396
\end{thebibliography}
\end{document}